\documentclass[12pt]{article}

\usepackage{graphicx}

\newcommand{\mre}{\mathrm{e}}
\newcommand{\mrO}{\mathrm{O}}
\newcommand{\mrd}{\mathrm{d}}

\newcommand{\Z}{{\mathbb{Z}}}

\newcommand{\order}[1]{\mathrm{O}\left( #1 \right) }

\newcommand{\FV}{\mathrm{FV}}

\newcommand{\gbar}{\overline{g}}

\newcommand{\ms}{\mathrm{{MS\kern-0.14em}\kern0.14em}}
\newcommand{\msbar}{\mathrm{\overline{MS\kern-0.14em}\kern0.14em}}
\newcommand{\SUN}{\mathrm{SU}(N)}
\newcommand{\SUtw}{\mathrm{SU}(2)}
\newcommand{\SUth}{\mathrm{SU}(3)}
\newcommand{\SSUN}{\SUN\times\SUN}
\newcommand{\SSUtw}{\SUtw\times\SUtw}
\newcommand{\SSUth}{\SUth\times\SUth}
\newcommand{\SOf}{\mathrm{SO}(4)}
\newcommand{\Of}{\mathrm{O}(4)}

\newcommand{\J}{\mathrm{J}}

\newcommand{\Lt}{L_t}

\newcommand{\Tr}{{\rm Tr}\,}

\newcommand{\hh}{\hat{h}}

\usepackage{a4}
\usepackage{a4wide}

\usepackage{amsmath}
\usepackage{amscd}
\usepackage{amsfonts}
\usepackage{amssymb}
\usepackage{mathrsfs}
\usepackage{fancybox} 
\usepackage{enumerate}
\usepackage{accents} % need to use \ring

\usepackage{bm}
\usepackage{amscd}
\usepackage{graphicx}
\usepackage[dvips]{color}
\usepackage{epsfig}

\makeatletter

\@addtoreset{equation}{section}
\makeatother

\usepackage{amsthm} % This is for \newtheorem*
\makeatletter

\@addtoreset{equation}{section}
\makeatother

\begin{document}

\begin{titlepage}
\title{
\vspace{-2cm}
\begin{flushright}
\normalsize{ 
MPP-2019-226}
\end{flushright}
       \vspace{1.0cm}
{{\bf On the rotator Hamiltonian for the SU}$\bold{(N)\,\times\,}${\bf SU}$\bold{(N)}$
{\bf sigma-model in the delta-regime}}
\vspace{0.7cm}
}
\author{
J.\ Balog\thanks{balog.janos@wigner.mta.hu},\;
F.\ Niedermayer${}^{\natural}$,\;
P.\ Weisz\thanks{pew@mpp.mpg.de}
\\[25pt]
${}^*$ {\it Institute for Particle and Nuclear Physics,}\\ 
{\it Wigner Research Centre for Physics,}\\ 
{\it MTA Lend\"ulet Holographic QFT Group,}\\
{\it 1525 Budapest 114, P.O.B.\ 49, Hungary}
\\[10pt]
${}^{\natural}$ {\it Albert Einstein Center for Fundamental Physics,} \\
{\it Institute for Theoretical Physics, University of Bern, Switzerland}
\\[10pt]
${}^{\dagger}$ {\it Max-Planck-Institut f\"ur Physik,} \\
{\it 80805 Munich, Germany}
\\[10pt]
}

\date{}
\maketitle

\thispagestyle{empty}

% \vspace{.2cm}

\begin{abstract}
\vspace{0.3cm}
\normalsize
We investigate some properties of the standard rotator
approximation of the $\SSUN$ sigma-model in the delta-regime. 
In particular we show that the isospin susceptibility calculated in this 
framework agrees with that computed by chiral perturbation theory up to
next-to-next to leading order in the limit $\ell=L_t/L\to\infty\,.$
The difference between the results involves terms vanishing 
like $1/\ell\,,$ plus terms vanishing exponentially with $\ell\,$. 
As we have previously shown for the O($n$) model, 
this deviation can be described by a correction to the rotator spectrum 
proportional to the square of the quadratic Casimir invariant. 
Here we confront this expectation with analytic 
nonperturbative results on the spectrum in 2 dimensions for $N=3\,.$
\end{abstract}
\end{titlepage}
%\tableofcontents

\section{Introduction}
\label{Intro}

The low energy dynamics of QCD in the $\delta-$regime
is to lowest order in chiral perturbation theory ($\chi$PT) 
described by a quantum rotator for the spatially
constant Goldstone modes \cite{Leutwyler:1987ak}.
We recall that for a system in a periodic spatial box 
of sides $L$ the $\delta-$regime is where the ``time" extent $\Lt \gg L$
and $m_\pi L$ is small (i.e. small or zero quark mass) 
whereas $F_\pi L$, ($F_\pi$ the pion decay constant) is large.

Many other systems described by non-linear sigma models, 
also in $d=2,3$ dimensions, 
are similarly approximated by a quantum rotator 
to leading order in the analogous perturbative domain.
Accordingly, the lowest energy momentum zero states 
in a representation $r$ of the symmetry group have, 
to leading order perturbation theory, energies of the form 
\begin{equation}\label{E_Casimir}
E(r) \propto\mathcal{C}_2(r)\,,
\end{equation}
where $\mathcal{C}_2(r)$ 
is the eigenvalue of the quadratic Casimir (of the symmetry group)
in the representation $r\,$.

At 1-loop level it turns out that the Casimir scaling \eqref{E_Casimir}
still holds, but it is of course expected that at some higher order 
the standard rotator spectrum will be modified. 
The standard rotator describes a system where the length of the total
magnetization on a time-slice does not change in time. 
This is obviously not true in the full effective model given by $\chi$PT.

In a previous paper \cite{SUNxSUN}
we pointed out that by comparing the already obtained 
NNLO results for the isospin susceptibility 
from $\chi$PT at large $\ell\equiv L_t/L$ 
with that computed from the standard rotator,
one can establish, under reasonable assumptions,
that at 3-loops there is a correction to the rotator Hamiltonian
proportional to the square of the Casimir operator,
with a proportionality constant determined by the NNLO
low energy constants (LEC's) of $\chi$PT. 

In ref.~\cite{chi_rot_On} we considered the QM rotator 
for the group O($n$). In this paper we extend the analysis 
of the QM rotator to the group $\SSUN\,$, 
which has for $N>2\,$, to our knowledge,
not been frequently considered in the literature.

This paper is organized as follows: In section 2 we recall the definition
and results for the isospin susceptibility of the standard quantum rotator
coupled to a chemical potential for the SU$(2)\times {\rm SU}(2)$ and the
SU$(3)\times {\rm SU}(3)$ cases. Results for the general 
SU$(N)\times {\rm SU}(N)$ case are given in section 3. 
The results in this section are new, in particular
Eq. (\ref{SUNconj}) gives the eigenvalue of the quadratic Casimir invariant for
a generic SU$(N)$ representation. In section 4 we discuss the corrections to the
simple rotator formula calculated in chiral perturbation theory.
In sect.~\ref{cased2} we consider the case of $d=2\,$. 
For $\SSUth$ Kazakov and Leurent \cite{Kazakov} have computed the 
lowest energies of two representations using an alternative to
the thermodynamic Bethe ansatz (TBA). Their NLIE (nonlinear integral equation),
in contrast to the infinite component TBA, is formulated in terms of finitely
many unknown functions and allows for a much better numerical precision than
the corresponding TBA calculation. Their data clearly show that 
Casimir scaling is valid to a very good approximation for $ML<1\,,$ 
however it was not sufficiently precise to see the expected deviations.
Here we present more precise numerical data allowing us to clearly see
the deviation from the simple rotator spectrum. Our data are completely
consistent with the results of the perturbative calculations.
The details of our calculations are given in various appendices, in particular
the algorithm of Ref. \cite{Kazakov} to use the NLIE equations 
for the calculation
of the finite size spectrum of the model (for $N=3$) is reviewed in appendix E.

The contribution of the main author of this paper, Ferenc Niedermayer, 
was essential in the formulation of the bulk of this paper. 
His untimely death on 12 August 2018 denied him the completion of the 
numerical calculations. 
We devote this paper to the memory of Ferenc.

\section{The isospin susceptibility}
\label{susceptibility}

Here we consider the Hamiltonian of the $\SSUN$ standard quantum rotator 
with a chemical potential coupled to generators $J_{L3}\,,J_{R3}$:
\begin{equation}
  H_0(h) = \frac{1}{\Theta}\left[J_L^2  + J_R^2 \right]
                         +h\left[J_{L3} - J_{R3}\right]\,, 
  \label{H0h} 
\end{equation}
where $J_X^2$ are the quadratic Casimir operators of 
the left and right $\SUN$ groups: 
\begin{equation}
J_X^2=\sum_{i=1}^{N^2-1}J_{Xi}^2\,,\,\,\,\,X=L,R\,,
\end{equation}
and $\Theta$ is the moment of inertia. 
In $d=4$ dimensions
to lowest order $\chi$PT one has $\Theta\simeq F^2 L^3\,$.

The isospin susceptibility is defined as the second derivative 
of the free energy wrt $h$:
\begin{equation}
  \chi= \left.\frac{1}{\Lt L^{d-1}}
  \frac{\partial^2}{\partial h^2}\ln Z(h;u)\right|_{h=0}\,, 
  \quad\quad Z(h;u)=\Tr \exp\{-H_0(h)L_t\}\,,
\end{equation}
where $u=2 L_t/\Theta\,.$
The partition function has for small $h\,,$ the expansion
\begin{equation} \label{Zuh}
  Z(u;h) = z_0(u) + \frac12 h^2 L_t^2 z_1(u) + \order{h^4} \,,
\end{equation}
with
\begin{align}\label{z0x}
z_0&=\Tr\exp\{-H_0(0)L_t\}\,,
\\ \label{z1x}
z_1&=\Tr\left[\left(J_{L3} - J_{R3}\right)^2\exp\{-H_0(0)L_t\}\right]\,.
\end{align}
The isospin susceptibility is then given by
\begin{equation} 
  \chi=\frac{L_t}{L^{d-1}}\frac{z_1(u)}{z_0(u)}\,.
\end{equation}
We wish to compute $\chi$ for small $u$ for general $N\,,$
however the reader may find it instructive to first consider 
the special cases $N=2,3$ which we treat in the following subsections. 

\subsection{$\SSUtw$ case}

The quantum mechanics (QM) of a symmetric rotor (rigid body) in 3 dimensions 
is equivalent to QM of a point particle moving in the SU(2) group manifold,
which is the sphere $\mathrm{S}_3\,$. It can be considered as a special
case of the O($n$) rotator (point particle moving on the sphere 
$\mathrm{S}_{n-1}$) for $n=4\,$.
At the same time it is a special case of a particle moving 
on the $\SUN$ group manifold with $N=2\,$.

The coordinates in the two descriptions are:
$U\in \SUtw$, where $U=s_0 + i s_k \sigma_k\in\SUtw\,$, 
($\sigma_k$ the Pauli matrices)
or equivalently, in the O(4) picture
$\mathbf s = (s_0, s_1, s_2,s_3) \in \mathrm{S}_3$,  ($\mathbf s^2=1$)\,.
The wave functions have the form  $\psi(U)$ or $\psi(\mathbf s)\,$.
The symmetry group of $H_0$ for $h=0$ is
$G=\SSUtw \simeq \SOf\,,$ and the
transformation of a wave function under $g=g_L\times g_R\in G$:
$$\psi(U) \to \psi(g_L^{-1} U g_R)\,,\,\,\, {\rm or}\,\,\,
\psi(\mathbf s) \to \psi(O_g^{-1}\mathbf s)\,.$$
The symmetry generators are
$J_{Li}$ for $\SUtw_L$ and $\mathbf J_{Ri}$ 
for $\SUtw_R$ transformations, ($i=1,2,3$),
or alternatively the 6 generators of $\SOf\,$.

In the $\SSUtw$ picture the wave functions are constructed using 
the $U$ variables. The set of four wave functions 
$\psi(U) \in {U_{11}, U_{12}, U_{21}, U_{22} }$ belong to the representation
with $j_L=j_R=1/2$. 
In general, the Hilbert space of the system splits into 
irreps of $\SSUtw$ for which $j_L=j_R=j\,$
\footnote{In the classical description a given trajectory $U(t)$ 
of the particle can be reached in two equivalent ways, by left rotations 
$U(t)=g_L(t)U_0$ or by right rotations, $U(t)=U_0 g_R^\dagger(t)$.
Obviously, the energy of a given eigenstate or its multiplicity should not
depend on the description chosen.}.

To label the irreps of $\SUtw$ we adopt the convention which is
a special case to be used for $\SUN$ with $N\ge 3$ below.
The representation with given $j$ is denoted by $(p)$
where $p=2j=0,1,2,\ldots$, with the corresponding dimension $p+1=2j+1\,$.
Accordingly, the eigenstates of the Hamiltonian \eqref{H0h} 
$\vert j,m_L\rangle \times \vert j,m_R\rangle$, $-j\le m_L,m_R \le j\,$
belong to the representation $(p)\times(p)$ with multiplicity $(p+1)^2\,$
\footnote{Note that in the equivalent O(4) language the eigenstates with given 
$l=0,1,2,\ldots$ have multiplicity $(l+1)^2$ (cf. \cite{chi_rot_On}).}.

%\begin{equation} \label{Q0_SU2}
%  \left[Q_0(p)\right]^2=(p+1)^2 \,.
%\end{equation}

The eigenvalue of the quadratic Casimir invariant in a representation
$(p)$ is given by
\begin{equation} \label{C_SU2}
  C_2^{(2)}((p))=\frac14 p(p+2) = \frac14 (p+1)^2 -\frac14\,, 
%= \tilde{C}_2(p)-\frac14 \,.
\end{equation}
%Note that apart from the constant the Casimir is purely quadratic 
%in the variable $\tilde{p}=p+1$.
which differs by a factor 4 from the O(4) Casimir invariant $l(l+2)$
\footnote{Note that $\exp(i J_3 \phi)$  in $\SUtw$ rotates by an angle
$\phi/2$ around the 3rd axis,  while for O(4) $\exp(i J_3 \phi)$ 
rotates by angle $\phi$.}.
The kinetic energy is then given by
\begin{equation} \label{E_kin}
  E_{\text{kin}} = \frac{2 C_2^{(2)}((p))}{\Theta}
                 = \frac{C_{\Of}(l)}{2\Theta}\,,
\end{equation}
consistent with our conventions in ref.~\cite{chi_rot_On}. 

In the O(4) picture the isospin chemical potential 
is coupled to generator of rotations in the 12-plane, $L_{12}\,$. 
It has eigenvalues $m=-l,\ldots,l$ for $\SOf$. 
The corresponding multiplicities are $g_{lm}=l-|m|+1\,$.
For $l=1$ one has: $m=\pm 1$: $s_1\pm i s_2$, 
for $m=0$:  $\{s_0, s_3 \}$.
In the $\SSUtw$ picture in the $(1)\times(1)$ irrep 
$m=1$ corresponds to wave function $U_{21}=s_1+is_2$, 
$m=-1$ to $U_{12}=s_1-is_2$,
while $m=0$ to wave functions $U_{11}$ and $U_{22}\,$.

The partition function with zero chemical potential is then
\begin{equation} \label{z0_SU2}
  z_0(u) = \sum_{p=0}^\infty \mre^{-u C_2^{(2)}((p))} (p+1)^2
  = \frac12 \mre^{u/4}\sum_{k=-\infty}^\infty \mre^{-u k^2/4} k^2
  = -\frac{1}{2\pi}\mre^{u/4}S'\left( \frac{u}{4\pi}\right)\,,
\end{equation}
where $S(x)$ is the Jacobi theta-function 
\begin{equation}
S(x)= \sum_{n=-\infty}^\infty \mre^{-\pi x n^2}\,.
\end{equation}
Using $S(x)=x^{-1/2}S(1/x)$ one obtains
\begin{equation} \label{z0_SU2_A}
  z_0(u) = \sqrt{4\pi}\mre^{u/4}u^{-3/2}S\left(\frac{4\pi}{u} \right)
  +16 \pi^{3/2}\mre^{u/4}u^{-5/2} S'\left(\frac{4\pi}{u} \right)\,.
\end{equation}
For small $u$ it has an expansion
\begin{equation} \label{z0_SU2_C}
  z_0(u) = 2\sqrt{\pi}\mre^{u/4} u^{-3/2} +
  \order{u^{-5/2} \mre^{-4\pi^2/u}} \,.
\end{equation}

For $z_1$ we have
\begin{equation} \label{z1_SU2}
\begin{split}
  z_1(u) &= 2\sum_{p=0}^\infty \mre^{-u C_2^{(2)}((p))}(p+1)
  \sum_{s=0}^p(p/2-s)^2
\\
  &= \frac{1}{12}\mre^{u/4}\sum_{k=-\infty}^\infty\mre^{-u k^2/4}k^2(k^2-1)
   = -\frac23 \frac{\partial z_0(u)}{\partial u}\,.
\end{split}
\end{equation}
The $\SSUtw$ rotator susceptibility is then given by
\begin{equation} \label{chi_SU2_rot}
  L^{d-2}\chi_{\text{rot}} = \frac{\Theta}{2 L} - \frac{\ell}{6} + \ldots
\end{equation}
where $\ell=L_t/L$ with no power-like corrections!
This is in agreement with the O($n$) rotator result (2.3) 
of \cite{chi_rot_On} at $n=4$ and with the $\chi$PT 
for $\SSUN$ at $N=2$ \cite{SUNxSUN}.

\subsection{\boldmath The $\SSUth$ case} 

Next we consider the QM of a point particle moving on the group manifold of 
$\SUth$. For the $\SUth$ irreducible representations we shall 
in this subsection use the familiar notation 
$(p,q)$ where $p,q=0,1,2\ldots\,$ i.e. the first and second rows
of the corresponding Young tableaux have $p+q$ and $q$ boxes respectively.

%It is convenient to introduce the notation $\tilde{p}=p+1$, $\tilde{q}=q+1$.
The corresponding value of the quadratic Casimir invariant is
\begin{equation} \label{C2_SU3}
  C_2^{(3)}((p,q)) = \frac13 \left( p^2+ q^2+pq+3p+3q \right)\,,
%  = \frac13 \left( \tilde{p}^2+\tilde{q}^2+\tilde{p}\tilde{q}\right) - 1\,,
  \end{equation}
while the dimension of the representation is given by
\begin{equation} \label{dpq}
  d(p,q) = \frac12 (p+1)(q+1)(p+q+2)\,.
%  = \frac12 \tilde{p}\tilde{q}(\tilde{p}+\tilde{q})\,.
\end{equation}

We consider a system described by coordinates
$U\in \SUth$, and wave functions $\psi(U,U^\star)$  which 
transform under $g=g_L\times g_R\in \SSUth$ according to
$$\psi(U,U^\star) \to \psi(g_L^{-1} U g_R\,, g_L^T U^\star g_R^\star)\,.$$ 
The 9 wave functions $\psi(U) = U_{ab}$, where $a,b\in\{ 1,2,3 \}$
belong to the representation $(1,0)\times(0,1)$ of $\SSUth$.
The first index, $a$ is for $\mathbf 3\equiv (1,0)$, 
while $b$ for $\overline{\mathbf 3} \equiv (0,1)$.

At this stage we assume that the irreps appearing in the partition function
sum over states are of the type $(p,q)\times (q,p)\,$;
the motivation for this will be given in subsection~\ref{wavefns}
\footnote{From products of $n$ matrix elements $U_{ab}$
one finds the irreps $(p,q)\times(q,p)$ with
$p=n-2k$, $q=k$, for $k=0,1,2,\ldots,k_{\mathrm{max}}$,
where $k_{\mathrm{max}}=n/2$ for even $n$, and $k_{\mathrm{max}}=(n-1)/2$
for odd $n$.}.
The corresponding energy is given by (cf. \eqref{E_kin})
\begin{equation} \label{E_kin3}
  E_{\text{kin}}=\frac{2 C_2^{(3)}((p,q))}{\Theta}\,,
\end{equation}
with the multiplicity $d(p,q)^2$ (cf. \eqref{dpq}).

We discuss the full dependence of $Z(u;h)$ on $h$ in Appendix \ref{Zuh_SU3}
although this information will not be needed in this paper.
From eqs.~\eqref{z0x},\eqref{z1x} we obtain
\begin{equation}
\begin{aligned}\label{z0_SU3}
  z_0(u) &= \sum_{p,q=0}^{\infty}\mre^{-u C_2^{(3)}((p,q))} 
            Q_0^{(3)}((p,q))^2 \,,
\\
  z_1(u) &= \sum_{p,q=0}^{\infty}\mre^{-u C_2^{(3)}((p,q))} 
          2 Q_0^{(3)}((p,q)) Q_2^{(3)}((p,q))\,,
\end{aligned}
\end{equation}
with 
\begin{equation}\label{Qk_SU3}
Q_k^{(3)}((p,q))\equiv\sum_{s\in(p.q)}\lambda(s)^k\,,
\end{equation}
where $s$ are eigenstates of $J_3$ with eigenvalues $\lambda(s)$.
One has (see \eqref{C1Q2})
\begin{align}
  Q_0^{(3)}((p,q)) &= d(p,q) \,, \label{Q0} \\
  Q_2^{(3)}((p,q)) &= \frac18 C_2((p,q)) d(p,q) \,. \label{Q2}
\end{align}

In $z_0,z_1$ we have a double sum over integers, 
hence analytic expressions are not so simple. 
However the leading terms for small $u$  
can be determined analytically. After separating
the constant term in \eqref{C2_SU3}, the remaining expressions
are homogeneous in $\tilde{p}=p+1$ and $\tilde{q}=q+1\,$:
\begin{equation}
  z_0(u) = \frac14\mre^{u}\sum_{\tilde{p},\tilde{q}=1}^{\infty}
            \exp\left[-\frac{u}{3}\left(\tilde{p}^2+\tilde{q}^2
             +\tilde{p}\tilde{q}\right)\right]
            \tilde{p}^2\tilde{q}^2(\tilde{p}+\tilde{q})^2\,.
\end{equation}
For small $u$ we can replace the sums by integrals to obtain
\begin{equation}\label{z0A}
  z_0(u)\sim z_{0A}(u) = A_0^{(3)} \mre^{u} u^{-4} \,,
\end{equation}
with
\begin{equation}
  A_0^{(3)}  = \frac14\int_0^{\infty}\mrd x\,\mrd y\,
            \exp\left[-\frac13\left(x^2+y^2+xy\right)\right]
            x^2y^2(x+y)^2=\pi \sqrt{3}\,.
\end{equation} 

To investigate the corrections to \eqref{z0A} one can first proceed 
numerically e.g. evaluating the difference to 500 digits at $0.1 \le u \le 1$
one has for the relative deviation $(z_0(u)-z_{0A}(u))/z_0(u)$
at $u=1.0$:  $\sim 10^{-13}$, and at $u=0.1$: $\sim 10^{-164}\,$,
i.e. it decreases faster than any power of $u$.
Fitting the difference one obtains the next approximation
\begin{equation}
  z_{0}(u) = \pi\sqrt{3}\, \mre^{u} u^{-4}
  - \sqrt{3}\left(\frac{2\pi}{u}\right)^7\mre^{-4\pi^2/u}
  \left(1+\order{u}\right) \,.
\end{equation}
The correction to the leading first term is exponentially small 
for small $u$, and has a structure similar to \eqref{z0_SU2_C}.

For $N=3$ using \eqref{z0_SU3} this gives for the susceptibility
of the $\SSUth$ rotator 
\begin{equation} \label{chi_SU3_rot}
  L^{d-2} \chi_{\text{rot}} = \frac{\Theta}{2 L} -\frac14 \ell + \ldots \,.
\end{equation}
We stress again that for $u\to 0$ the omitted terms decrease faster than 
any power of $u$.
The leading term in \eqref{chi_SU2_rot}, \eqref{chi_SU3_rot}  
is the classical result for the high temperature expansion of
the corresponding rotator (rigid body). 
The next one, $\propto \ell$ is the leading quantum correction,
which does not depend on $\Theta$, only on the corresponding group.
It is interesting to note that for $N=2,3$ the $1/(F^2 L^2)$ term (for $d=4$)  
is absent in the expansion, a property which we will see
holds for arbitrary $N$.

\section{\boldmath The isospin susceptibility for general $\SSUN$}

In this section we extend the considerations in the last two subsections
for $N=2,3$ to general $N\,.$

\subsection{The quadratic Casimir invariant}

As proposed by Gelfand and Tsetlin \cite{Gelfand},
an irrep of $\SUN$ can be conveniently described
by a non-increasing series of $N$ integers 
(cf. \cite{SUN_CG} and references therein)
$m_1 \ge m_2 \ge \ldots \ge m_N\,$.
Two series differing in a constant, $m'_k = m_k + c\,,\,\forall k$ 
where $c\in \mathbb{Z}$ describe the same irrep.
One can choose $m_N=0$, however, for some purposes
it is convenient to use the redundant form with $N$ integers.

If one sets $m_N=0$ then $m_k$ corresponds to the number of
boxes in the $k$'th row of the corresponding Young tableau.
The more conventional description of an $\SUN$ irrep,
like $(p,q)$ for $\SUth\,$, is given by the differences 
$(p_1,p_2,\ldots,p_{N-1})$ where $p_k=m_k-m_{k+1} \ge 0\,$.

Following the notation in \cite{SUN_CG}, let $J_z^{(l)}\,,\,l=1,\dots N-1$ 
be a basis of the Cartan subalgebra. Together with generators
$J_\pm^{(l)}$ they generate SU(2) subalgebras for each $l$. The $J_z^{(l)}$
are normalized to have half-integer eigenvalues, and we can identify
$J_3$ with one of them, say $J_z^{(1)}$. In a given representation
$r$ there is a highest weight vector $\vert H\rangle$ which is annihilated by
all $J_+^{(l)}$. Its eigenvalues are given by 
$J_z^{(l)}\vert H\rangle=\lambda_l(H)\vert H\rangle$
with $\lambda_l(H)=p_l/2\,.$ 

Eq.~\eqref{Qk_SU3} is generalized to
\begin{equation}\label{Qk_SUN}
Q_k^{(N)}(r)\equiv\sum_{s\in r}\lambda_1(s)^k\,,
\end{equation}
where $s$ are eigenstates of $J_3$ with eigenvalues $\lambda_1(s)$.
$Q_0^{(N)}(r)$ is the dimension of a given irrep $r$ and
is explicitly given by \cite{SUN_CG}
\begin{equation} \label{Q0_SUN}
  Q_0^{(N)}((m_1,\dots,m_N)) = 
  \prod_{1\le k < k' \le N}\left( 1 + \frac{m_k-m_{k'}}{k'-k}\right)\,. 
\end{equation}

The quadratic Casimir invariant can be calculated using 
the basis of the $\mathrm{su}(N)$ algebra described in \cite{SUN_CG}.
Alternatively one can use recursion relations 
for $Q_0^{(N)}$ and $Q_2^{(N)}\,,$ discussed in appendix
~\ref{app_recursion} to obtain $C_2^{(N)}$ using 
\begin{equation} \label{C1Q2}
  \frac{Q_2^{(N)}(r)}{Q_0^{(N)}(r)} = \langle J_3^2 \rangle_r = 
  \frac{1}{N^2-1} \langle J^2 \rangle_r
  = \frac{1}{N^2-1} C_2^{(N)}(r)\,,
\end{equation}
where $N^2-1=\text{dim}(\SUN)$ is the dimension of the group.
The recursion relations from $Q_s^{(N-1)}$ to $Q_s^{(N)}$, 
($s=0,2$) contain $N-1$ nested summations. 
For not too large $N$ one can perform these analytically.
We have done this for $N\le 5\,$, 
and obtained a very simple result, 
which is easy to generalize to arbitrary $N\,$.
%Introducing the sums
%\begin{equation} \label{MsN}
%    M_r^{(N)} = \sum_{k=1}^N m_k^r \,, 
%\end{equation}
We conjecture
\begin{equation}  
   C_2^{(N)}((m_1,\dots,m_N)) = \frac12 \sum_{k=1}^N m_k^2 
      - \frac{1}{2N}\left(\sum_{k=1}^N m_k\right)^2 
    + \sum_{k=1}^N \left( \frac{N+1}{2}-k\right) m_k \,.
\label{SUNconj}
\end{equation}
Note that this expression is invariant under a constant shift
$m_k \to m_k + c\,$ as it should.

Denoting $n_k=m_k+N-k$ one has for the factor appearing in $Q_0^{(N)}$
in \eqref{Q0_SUN},
\begin{equation}
  1+ \frac{m_k-m_{k'}}{k'-k} = \frac{1}{k'-k} (n_k-n_{k'})\,.
\end{equation}
Hence
\begin{equation}
  Q_0^{(N)}((m_1,\ldots,m_N)) = 
  \left. \overline{Q}_0^{(N)}(n_1,\ldots,n_N)\right|_{n_k=m_k+N-k}
\end{equation}
where
\begin{equation}\label{QovN}
  \overline{Q}_0^{(N)}(n_1,\ldots,n_N)
   = \mathcal{B}_N \prod_{1\le k < k' \le N} (n_k-n_{k'})
\end{equation}
with 
\begin{equation}
  \frac{1}{\mathcal{B}_N} = \prod_{1\le k < k' \le N} (k'-k) 
   = 2!\,3!\ldots (N-1)! \,.
\end{equation}

% However, because of the factors $(n_k-n_{k+1})$
% appearing in $Q_0$, for the partition function one can extend 
% the range of summation to $n_k \ge n_{k+1}$.

The Casimir invariant of the representation in terms of $n$'s is
\begin{equation} \label{CovN}
  \begin{aligned}
    \overline{C}_2^{(N)}(n_1,\ldots,n_N) & = \frac12 \sum_{k=1}^N n_k^2 
    - \frac{1}{2N}\left( \sum_{k=1}^N n_k \right)^2 -c_N
    \\
    & = \frac{1}{2N} \sum_{1\le k < k' \le N} (n_k - n_{k'})^2 - c_N \,,
  \end{aligned}
\end{equation}
where 
\begin{equation}  
c_N \equiv \frac{N (N^2-1)}{24}
\end{equation}
is proportional to the curvature of the SU$(N)$ manifold.
Note that apart from the constant in $\overline{C}_2^{(N)}$ 
both expressions, \eqref{QovN} and \eqref{CovN},
are homogeneous in the new variables.

\subsection{Wave functions}
\label{wavefns}

First we note that for  $U \in \SUN$ the complex conjugate of a 
matrix element equals the corresponding cofactor of the matrix, 
\begin{equation}
  U^\star_{ab} = (-1)^{a+b} 
  \det\left( \left( U_{ij} \right)_{i\ne a\,, j\ne b} \right) \,.
\end{equation}
As a consequence, a function  written in terms of products
containing $U$'s and $U^\star$'s can be written in terms
of the $U$'s alone.

Under a general $\SSUN$ transformation one has
\begin{equation} 
  U \to U' = g_L^{-1} U g_R)\,.
\end{equation}
Under separate left/right transformations
\begin{equation} 
      (g_L^{-1}U)_{a a'} = (g_L^{-1})_{ab} U_{b a'}\,, 
\quad (Ug_R)_{a a'} = (g_R^{-1})^\star_{a' b'} U_{a b'}\,,
\end{equation}
i.e.\ $U$ belongs to the representation
$(1,0,\ldots,0) \times (0,\ldots,0,1)$,
according to its 1st and 2nd index, respectively.

Similarly for an arbitrary  representation $(r)$
\begin{equation} 
  \begin{aligned} 
    \left[D^{(r)}(g_L^{-1} U)\right]_{i i'} & =  
    \left[D^{(r)}(g_L^{-1})\right]_{ij}  \left[D^{(r)}(U)\right]_{j i'}\,,
    \\
    \left[D^{(r)}(U g_R)\right]_{i i'} & =  
    \left[D^{(r)}(g_R^{-1})\right]^\star_{i'j'}\left[D^{(r)}(U)\right]_{i j'}\,,
  \end{aligned} 
\end{equation}
where $1 \le i,i',j,j'\le \dim(r)\,$.
Hence the elements of the matrix $D^{(r)}(U)$ belong to a 
representation with complex conjugate pair $(r)\times(r^*)$, 
i.e.\ $(p_1,p_2,\ldots,p_{N-1})\times (p_{N-1},\ldots,p_2,p_1)$.
Strictly one should still show also that each such representation 
enters only once in the Hilbert space of the $\SSUN$ rotator;
here we accept this as a reasonable hypothesis.

\subsection{The partition function and susceptibility}

The partition function is given by (set $m_N=0$)
\begin{equation}
    z_0^{(N)}(u) =
    \sum_{m_1=0}^\infty
    \, \sum_{m_2=0}^{m_1}
    % \,\sum_{m_3=0}^{m_2}
    \ldots
    \,\sum_{m_{N-1}=0}^{m_{N-2}}
    \mre^{-u C_2^{(N)}((m_1,\ldots,m_{N-1},0))} 
    \left[Q_0^{(N)}((m_1,\ldots,m_{N-1},0))\right]^2 \,.
\end{equation}
% \begin{equation}
% %  \begin{aligned}
%     z_0^{(N)}(u) =
%     \sum_{n_1=N-1}^\infty
%     \, \sum_{n_2=N-2}^{n_1-1}
%     \,\sum_{n_3=N-3}^{n_2-1}
%     \ldots
%     \,\sum_{n_{N-1}=1}^{n_{N-2}-1}
%     \mre^{-u C_2^{(N)}(\ldots)} 
%     \left[Q_0^{(N)}(\ldots)\right]^2 \,,
%   %\end{aligned}
% \end{equation}
Changing to the variables $n_k=m_k+N-k$
the condition $m_k \ge m_{k+1}$ transforms into $n_k > n_{k+1}\,$. 
Also the irreps with $n'_k = n_k + c$ where $c\in \Z$ are equivalent
and should be taken only once in the partition function. 
Again a convenient choice is to set $n_N = 0$
% Alternatively, one can set $m=0$ in $\sum_k n_k=mN+\nu$,
% where $0 \le \nu \le N-1$.
and one has
\begin{equation} \label{z0Na0}
\begin{aligned}
    z_0^{(N)}(u)&= \mre^{u c_N} \sum_{n_1 > n_2 > \ldots > n_N} \delta_{n_N,0}
    \exp\left[ -\frac{u}{2} \sum_{k=1}^{N} n_k^2 
      + \frac{u}{2N} \left(\sum_{k=1}^N n_k \right)^2\right] 
    \overline{Q}_0^2(n_1,\ldots,n_N)\\
    &= \frac{\mre^{u c_N}}{N!} 
    \sum_{\{n\}=-\infty}^{\infty}  \delta_{n_N,0}
    \exp\left[ -\frac{u}{2} \sum_{k=1}^{N} n_k^2 
      + \frac{u}{2N} \left(\sum_{k=1}^{N} n_k \right)^2\right] 
    \overline{Q}_0^2(n_1,\ldots,n_N)\,,
\end{aligned}
\end{equation} 
where the second equality follows 
since the summand is invariant under permutations.

In the conventional ``p-notation'' the Casimir invariant
and $Q_0$ for the $r_p=(p,0,0,\ldots,0)$ representation is
\begin{equation} 
  \begin{aligned}
    C_2^{(N)}(r_p) &= \frac{(N-1)}{2N}p(p+N)  \,, \\
    Q_0^{(N)}(r_p) &= 
    \prod_{n=1}^{N-1}\left(1 + \frac{p}{n}\right)
    = \frac{(N+p-1)!}{p! (N-1)!} \,.
  \end{aligned}
\end{equation}
In particular for the ground state $p=0$
\begin{equation} 
  \begin{aligned}
    C_2^{(N)}(r_0) &= 0 \,, \\
    Q_0^{(N)}(r_0) &= 1 \,,
  \end{aligned}
\end{equation}
and for $p=1$
\begin{equation} 
  \begin{aligned}
    C_2^{(N)}(r_1) &= \frac{N^2-1}{2N} \,, \\
    Q_0^{(N)}(r_1) &= N \,.
  \end{aligned}
\end{equation}

For the adjoint representation\footnote{Here we assume $N\ge 3$}
$r_A=(1,0,\ldots,0,1)$ one obtains
\begin{equation} 
  \begin{aligned}
    C_2^{(N)}(r_A) &= N \,, \\
    Q_0^{(N)}(r_A) &= N^2-1 \,.
  \end{aligned}
\end{equation}

Since $C_2^{(N)}(r_1) < C_2^{(N)}(r_A)<C_2^{(N)}(r_2)$
the mass gap is given by the states in the representation
$r_1\times (0,\ldots,0,1)$, and its conjugate,
with a total multiplicity $2 N^2$.
The formula for the mass gap is then (cf. \cite{SUNxSUN} eq.~(4.31))
\begin{equation}  
  E_1 = \frac{N^2-1}{N \Theta}\,.
\end{equation}
The contribution of these states together with the ground state gives
\begin{equation}  
  z_0^{(N)}(u) = 1 + 2 N^2 \exp\left(-\frac{N^2-1}{2N} u\right)
  + (N^2-1)^2 \exp(-N u) + \ldots\,,\quad \text{for }  u\gg 1 \,.
\end{equation}

The behavior of $z_0^{(N)}(u)$ for $u\to 0$
is derived in Appendix~\ref{app_z0_u_small} with the result
\begin{equation} \label{z0NN}
    z_0^{(N)}(u) = A_N u^{-(N^2-1)/2} \mre^{uc_N} 
    \left[ 1 + \order{\mre^{-4\pi^2/u} u^{-2N+3}}\right] \,,  
\end{equation}
where
\begin{equation}
\begin{aligned}
 A_N &= \frac{\mathcal{B}_N^2}{(N-1)!\sqrt{2\pi N}}
  \int_{-\infty}^\infty \prod_{j=1}^N\left[\mrd n_j 
  \, \exp\left(-\frac12 n_j^2 \right)\right]\,
  \prod_{1\le k < k' \le N} (n_k-n_{k'})^2 
\\
 &=(2\pi)^{(N-1)/2}\sqrt{N}\mathcal{B}_N\,.
\end{aligned}
\end{equation}

Due to \eqref{C1Q2} we have (for general $N$)
\begin{equation} \label{z1z0_SUN}
  z_1(u) = -\frac{2}{N^2-1} \frac{\partial z_0(u)}{\partial u}\,.
\end{equation}
Thus the susceptibility is obtained from $z_0(u)$ as
\begin{equation} \label{chi_SUN_z0}
  L^2 \chi_{\text{rot}}
  =\frac{1}{L_t L} \frac{\partial^2}{\partial h^2}\ln Z(u;h)
  = \ell \frac{z_1(u)}{z_0(u)}
  =-\frac{2}{N^2-1}\ell\frac{\partial}{\partial u}\log(z_0(u)) \,.
\end{equation}

The susceptibility is then given by
\begin{equation} \label{chi_SUN_PT}
  L^2 \chi_{\text{rot}} =  \frac{\Theta}{2 L} 
  -\frac{N}{12} \ell 
  + \mrO\left(\frac{\ell^3}{F^4 L^4}\right)\,,\,\,
\end{equation}
which is in agreement with eq.~(4.48) of \cite{SUNxSUN} obtained
by $\chi$PT to NNL order for general $N\,$.

\section{\boldmath The $1/\ell$ term in $\chi$PT}

As mentioned in \cite{SUNxSUN} the susceptibility calculated 
in $\chi$PT to NNLO for $\ell\to\infty$ approaches the result obtained
in the rotator approximation.
However, the approach is not exponentially fast, to this order
one obtains besides the exponentially vanishing contribution 
a $\propto 1/\ell$ term (but no $\ell^{-k}$, $k\ge 2$ terms!).
More precisely for the deviation $\delta\chi = \chi - \chi_{\text{rot}}$ 
in \cite{SUNxSUN} (cf.\ eq.~(4.52)) we found
\begin{equation} \label{dchi_chi}
  \frac{\delta\chi}{\chi} \sim  \frac{c}{(F L)^4} \frac{1}{\ell}\,,\quad\quad
{\rm for}\,\,d=4\,.
\end{equation}
From here one concludes that the rotator spectrum should be distorted
at some higher order in $g_0^2$ already at small energies,
not only the energies $\sim L^{-1}$ of the $\mathbf{p}\ne 0$ modes.

Let us assume that the distortion of the spectrum has the form
\begin{equation} 
  \delta E(r) = \frac{\Phi(L)}{L}(C_2^{(N)}(r))^\kappa\,,
\end{equation}
then one obtains
\begin{equation} 
  \begin{aligned} 
    z_1(u) & = \frac{2}{N^2-1} 
    \left(-\frac{\partial }{\partial u}\right) z_0(u) \,,
    \\
    \delta z_0(u) & = - \Phi(L)\ell 
    \left(-\frac{\partial}{\partial u}\right)^\kappa  z_0(u) \,,
    \\
    \delta z_1(u) & = -\frac{2}{N^2-1} \Phi(L) \ell 
    \left(-\frac{\partial}{\partial u}\right)^{\kappa+1}  z_0(u) \,.
  \end{aligned} 
\end{equation}
Taking $z_0(u)\propto u^{-a}$ with $a=(N^2-1)/2$ one gets 
for the leading term
%\footnote{%
%For the leading correction in $\delta\chi/\chi$ one needs only the leading
%singularity $u^{-(N^2-1)/2}$ in \eqref{z0NN}.}
\begin{equation} 
  \begin{aligned} 
  \frac{\delta \chi_{\text{rot}}}{\chi_{\text{rot}}} 
  &= \frac{\delta z_1(u)}{z_1(u)} - \frac{\delta z_0(u)}{z_0(u)}
  = -\kappa\Phi(L) \ell u^{-\kappa} (a+1)(a+2)\ldots(a+\kappa-1)  + \ldots
  \\
  &= -2^{-\kappa} \kappa\Phi(L) \left(\frac{L}{\Theta}\right)^{-\kappa}
  \ell^{1-\kappa}(a+1)(a+2)\ldots(a+\kappa-1) + \ldots
  \end{aligned} 
\end{equation}
The observed deviation \eqref{dchi_chi} requires 
then $\kappa=2$ and since $\Theta\sim F^2L^3$ for $d=4$ 
we need $\Phi(L)\propto (FL)^{-8}$ for $d=4\,$. 

\section{\boldmath Delta regime in $d=2$}
\label{cased2}

The susceptibility computed in $\chi$PT is for $d=2$ given by
\cite{SUNxSUN}
\begin{equation}
  \chi=\frac{1}{2\gbar_\ms^2(1/L)}
   -\frac{N}{8\pi}\gamma_2^{(2)}(\ell)
   -\frac{N^2}{16\pi^2}r_2(\ell)\gbar_\ms^2(1/L)
    +\ldots
  \label{chi_d2}
\end{equation}
where $\gbar_\ms(q)$ is the minimal subtraction (MS) scheme running
coupling at momentum scale $q\,,$ and
\begin{equation}
  r_2(\ell)=\overline{w}(\ell)-2\kappa_{10}(\ell)-\frac12\gamma_2^{(2)}(\ell)
  \left(\alpha_1^{(2)}(\ell)-\frac{1}{\ell}-\frac12\gamma_2^{(2)}(\ell)\right)
  -\frac{1}{2\ell}\left(\gamma_3^{(2)}(\ell)+1\right)\,.
  \label{r2}
\end{equation}
The large $\ell$ behavior of the shape functions appearing in \eqref{chi_d2}
and \eqref{r2} are discussed in \cite{chi_rot_On}. 
In particular we find for $\ell\gg1$:
\begin{align}
\gamma_2^{(2)}(\ell)&\simeq -Z+\frac{2\pi\ell}{3}\,,
\\
r_2(\ell)&\simeq \frac34-\frac{Z}{2}-\frac{5\zeta(3)}{4\pi\ell}\,,
\end{align}
where
\begin{equation} \label{Zdef}
Z\equiv \ln(4\pi)-\gamma\,,\quad\quad(\gamma=-\Gamma'(1))\,.
\end{equation}

On the other hand the susceptibility computed from the simple rotator 
is given by:
\begin{equation}\label{chirot_d2}
  \chi_{\mathrm{rot}} = \frac{1}{2 \gbar_{\FV}^2(L) }
  -\frac{N}{12}\ell  + \mrO(\gbar_{\FV}^4(L))\,, \quad (d=2)\,,
\end{equation}
where $\gbar_{\FV}$ is the LWW running coupling \cite{LWW} 
defined through the finite volume mass gap: 
 \begin{equation}\label{FVcoupling}
   \gbar_{\FV}^2(L)\equiv \frac{N}{N^2-1}L E_1(L)\,.
 \end{equation}
Its expansion in terms of the running coupling in the MS scheme
of dimensional regularization (DR) is given by 
 \begin{equation}\label{gFVgMS}
   \gbar_{\FV}^2(L) = \gbar_{\ms}^2(1/L) + c_1 \gbar_{\ms}^4(1/L)
   + c_2 \gbar_{\ms}^6(1/L) +\cdots
 \end{equation}
The first two coefficients are obtained using the methods of ref.~\cite{LWW}:
 \begin{align}
   c_1 &= -\frac{N}{4\pi}Z\,,
\\
   c_2 &= \frac{N^2}{16\pi^2}\left(Z^2-Z+\frac32\right)\,.
 \end{align}
Combining the results we arrive at 
\begin{equation}\label{chidiffPT_d2}
  \frac{\chi-\chi_{\mathrm{rot}}}{\chi}
  =\frac{5N^2}{32\pi^3\ell}\zeta(3)\gbar_\ms^4(1/L)+\dots
\end{equation}

On the other hand, from our considerations of the modified rotator
in the previous subsection we would expect
 \begin{equation}\label{chidiffrot_d2}
\frac{\chi-\chi_{\mathrm{rot}}}{\chi}
= -\frac{1}{4\ell}\Phi_3(N^2+1)\gbar_{\ms}^4(1/L)+\dots
\end{equation}
where $\Phi_3$ is the leading coefficient in the perturbative expansion
of $\Phi(L)$, assuming the expansion starts at order $\gbar_{\ms}^8$:
\begin{equation}
\Phi(L)=\sum_{r=3}\Phi_r\gbar_{\ms}^{2r+2}(1/L)\,.
\end{equation}
Comparing \eqref{chidiffPT_d2} with \eqref{chidiffrot_d2} determines 
\begin{equation}
\Phi_3=-\frac{5N^2}{(N^2+1)\pi^3}f_3\,,
\end{equation}
where  
\begin{equation}
f_3=\frac18 \zeta(3)=0.15025711290\,.
\label{f3ex}
\end{equation}

The low-lying spectrum to order $\gbar_\ms^8$ is given by
\begin{equation}
  \begin{aligned}
    LE(r)&=2C_2^{(N)}(r)\gbar_\ms^2(1/L)
    \Big\{ 1+c_1\gbar_\ms^2(1/L)+c_2\gbar_\ms^4(1/L)
    \\
    & +\overline{c}_3 \gbar_\ms^6(1/L) + \ldots \Big\}
    + C_2^{(N)}(r)^2\Phi_3\gbar_\ms^8(1/L) 
    +\dots
  \end{aligned}
\end{equation}
where
\begin{equation}
  \overline{c}_3 = c_3 - \frac{(N^2-1)}{4N}\Phi_3\,.
\end{equation}

Hence we conclude, for example,
\begin{equation}
LE(r_1)-\frac{(N+1)}{2(N+2)}LE(r_2)
=-\frac{(N-1)^2(N+1)(N+3)}{4N^2}\Phi_3\,\gbar_\ms^8(1/L)+\dots\,.
\end{equation}
In subsection \ref{sectN3} we test this prediction for $N=3\,$. 

\subsection{Running coupling functions}

First, following Balog and Hegedus \cite{Balog:2003yr} we
introduce a function $\gbar_\J^2(L)$ of the box size $L$ through
\begin{equation}\label{gJcoupling}
\frac{1}{\gbar_\J^2(L)}+\frac{b_1}{b_0}\ln(b_0\gbar_\J^2(L))
=-b_0\ln(\Lambda_{\FV}L)\,, 
\end{equation}
where $b_0\,,b_1$ are the universal first perturbative coefficients of the
$\beta-$function
\footnote{The 3-loop coefficient in the MSDR scheme is 
$b_{2\ms}=3N^3/(64\pi^3)\,.$}
\footnote{Note $b_0,b_1$ in \eqref{b0b1} are factors $4,16$ respectively
larger than the coefficients $\beta_0,\beta_1$ given
in eq.(20) of ref.~\cite{Balog:1992cm}. The reason for this is that the
definition of the square of the  coupling in this  paper is a factor 
4 smaller than that in \cite{Balog:1992cm}.}:   
\begin{equation}\label{b0b1}
b_0=\frac{N}{2\pi}\,,
\quad\quad
b_1=\frac{N^2}{8\pi^2}\,,
\end{equation}
and $\Lambda_\FV$ is the $\Lambda-$parameter of 
the LWW finite volume coupling in \eqref{FVcoupling}. 
We chose the solution which is small for $\Lambda_{\FV}L$ small, 
which has the property
\footnote{For further remarks concerning this coupling see \cite{chi_rot_On}}
\begin{equation}
\gbar_\J^2(L)=\gbar_{\FV}^2(L)+\mrO(\gbar_{\FV}^6(L))\,,
\,\,\,\,\,\Lambda_{\FV}L\ll1\,.
\end{equation}
We consider $\gbar_J^2$ as a function of $z=ML$ where
$M$ is the infinite volume mass gap:
\begin{equation}
\frac{1}{\gbar_\J^2(z)}+\frac{b_1}{b_0}\ln(b_0\gbar_\J^2(z))
=-b_0\ln(z)+b_0\ln(M/\Lambda_{\FV})\,.
\end{equation}
The ratio $M/\Lambda_{\FV}$ is known using the result
in ref.~\cite{Balog:1992cm}
\footnote{
$M/\Lambda_\msbar=\sqrt{8\pi/\mre}\sin(\pi/N)/(\pi/N)$
and $\Lambda_\FV/\Lambda_\ms=\exp\left\{-Z/2\right\}=
\Lambda_\ms/\Lambda_\msbar\,.$}
\begin{equation}
\frac{M}{\Lambda_\FV}=
\sqrt{\frac{8}{\pi\mre}}N\mre^Z\sin\left(\frac{\pi}{N}\right)\,.
\end{equation}
Defining $\alpha_\J=\gbar_\J^2/(2\pi)$ the equation becomes 
\begin{equation}
\frac{1}{\alpha_\J(z)}+\frac{N}{2}\ln(\alpha_\J(z))=-N\ln(z)+J(N)\,,
\end{equation}
with
\begin{equation}
J(N)=\frac{N}{2}
\left[2Z+\ln\left\{\frac{N}{\pi}\sin^2\left(\frac{\pi}{N}\right)\right\}
-1+\ln(8)\right]\,.
\end{equation}

The LWW coupling has the following expansion in terms of $\gbar^2_\J$:
\begin{equation}
\gbar_{\FV}^2=\gbar^2_\J\left\{1+\frac{N^2}{2}\alpha_\J^2+\dots\right\}\,.
\end{equation}

\subsection{\boldmath Results 
for the $r=r_p=(p,0,0)\,,\,\,p=1,2$ energies for SU(3)} 
\label{sectN3}

In Table~\ref{tab:table1} we reproduce the data for the energy gaps $E(r_p)$
calculated from the numerical results given in Table \ref{tab:table2} in
appendix E.
$f_{3,\mathrm{est}}$ appearing in the last column is defined in \eqref{f3est}.
\begin{table}[h]
  \centering
  \caption{$\SSUth$ energies for representations 
$r_p=(p,0,0)\,,\,\,p=1,2$ }
  \label{tab:table1}
\vspace{0.5cm}
  \begin{tabular}{|l|l|l|l|l|l|}
    \hline
   \quad$z$ & $\quad\alpha_\J(z)$ &$\quad LE(r_1)$ & $\quad LE(r_2)$ 
   & $E(r_2)/E(r_1)$ & $f_{3,\mathrm{est}}$ \\
    \hline
    $0.01$&$0.03896665$&$0.6576493028$&$1.643478(3) $&$2.499019(4) $&$0.1856(9)$  \\
    $0.02$&$0.04264730$&$0.7208499394$&$1.8011779(6)$&$2.4986863(8)$&$0.1898(1)$  \\
    $0.03$&$0.04515478$&$0.7640825060$&$1.908986(4)$ &$2.498403(4) $&$0.1947(6)$  \\
    $0.04$&$0.04712791$&$0.7982109882$&$1.994061(4)$ &$2.498163(5) $&$0.1971(6)$  \\
    $0.05$&$0.04878634$&$0.8269748893$&$2.06573(1)$  &$2.49794(1)  $&$0.19936(15)$  \\
    $0.06$&$0.05023433$&$0.8521505407$&$2.1284405(6)$&$2.4977283(7)$&$0.20159(6) $  \\
    $0.07$&$0.05153033$&$0.8747342979$&$2.1846712(7)$&$2.4975255(8)$&$0.20357(7) $  \\
    $0.08$&$0.05271070$&$0.8953461736$&$2.2359752(2)$&$2.4973304(2)$&$0.20533(2) $  \\
    $0.09$&$0.05379972$&$0.9144006339$&$2.283386(3) $&$2.497140(3) $&$0.2070(2)  $  \\
    $0.1$ &$0.05481451$&$0.9321896996$&$2.327633(1) $&$2.496952(1) $&$0.20871(7) $  \\
    $0.2$ &$0.06264221$&$1.0705947394$&$2.671321(2) $&$2.4951754(2)$&$0.22245(2) $  \\
    $0.5$ &$0.07755223$&$1.3414119783$&$3.339568(1) $&$2.489591(1) $&$0.25595(2) $  \\
    $1.0$ &$0.09516460$&$1.6789366076$&$4.158326(2) $&$2.476762(2) $&$0.31546(2) $  \\
    \hline
  \end{tabular}
\end{table}
From the fifth column of Table~\ref{tab:table1}, we see
that the ratio $E(r_2)/E(r_1)$ is close to the ratio 10/4
of the Casimir eigenvalues. However, our numerical precision is sufficient
to establish that the simple effective rotator model requires corrections. 

\begin{figure}[htb]
  \centering
  \includegraphics[width=0.9\textwidth]{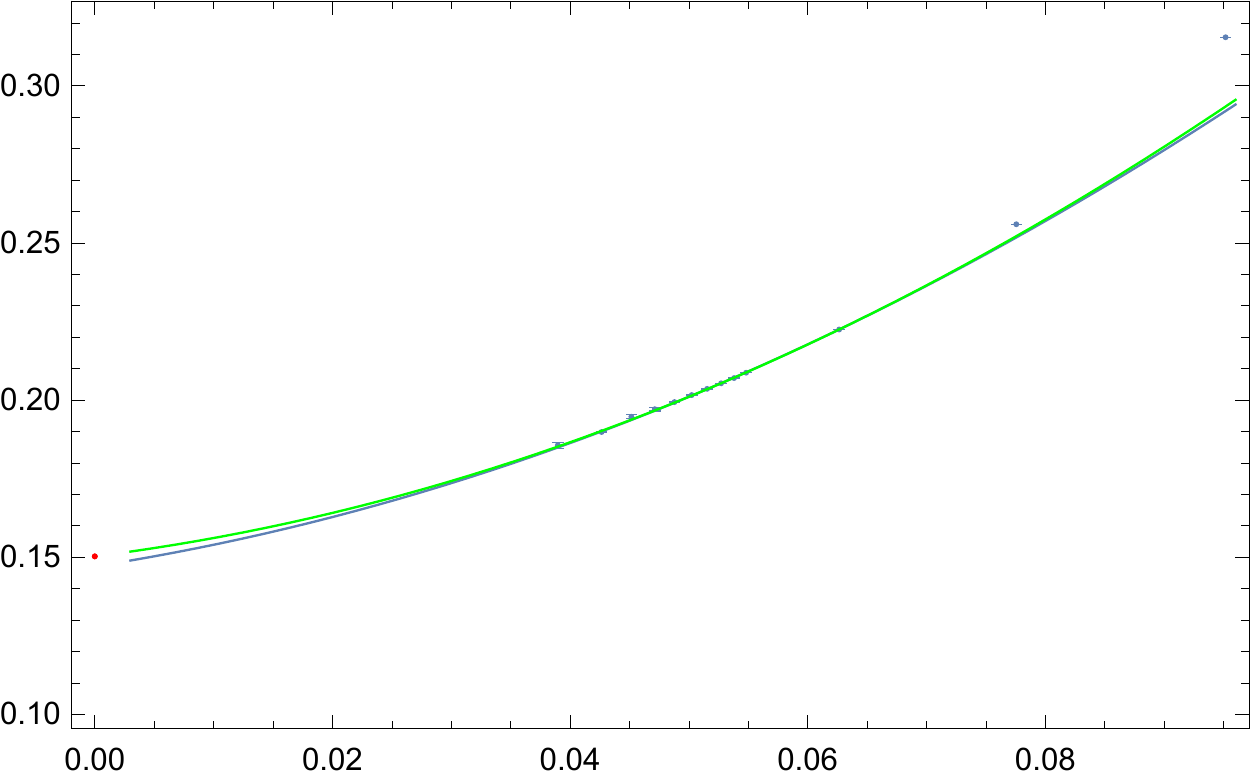}
  \caption{Plot of the estimate for $f_3$ given in \eqref{f3est}
    for SU(3); circles with error bars are data from Table \ref{tab:table1}.
    The blue line is a quadratic fit to the first 11 data points.
    The green line is a constrained quadratic fit where \eqref{f3ex} at
    $\alpha_{\rm J}=0$ is kept fixed.
}
  \label{fig_f3est_n3}
\end{figure}

To see even more clearly the agreement of our analysis with
the data in Table~\ref{tab:table1}
in Fig.~\ref{fig_f3est_n3} we plot estimates for $f_3$ given by
\begin{equation}\label{f3est}
f_{3,\mathrm{est}}=\frac{(N^2+1)}{10(N-1)^2(N+1)(N+3)2\pi\alpha_\J^4}
\left[LE(r_1)-\frac{(N+1)}{2(N+2)}LE(r_2)\right]\,,
\end{equation}
for the case $N=3$. In the figure the extrapolation to zero volume is also
shown. We do not have measured values close enough to  $\alpha_{\rm J}=0$ to make
linear fits. Our extrapolation is based on a quadratic least squares fit
(weighted by the error bars given in the last column in
Table~\ref{tab:table1}) to the first 11 data points in the range
$\alpha_\J < 0.063$ giving $f_3({\rm quadratic},\, 11)=0.147$. If we use the
first 10 data points only, we get $f_3({\rm quadratic},\, 10)=0.127$. The
green line is a constrained fit where the zero volume limit \eqref{f3ex} is kept
fixed  at $\alpha_{\rm J}=0$. This shows that our measurements are completely 
consistent with our prediction in \eqref{f3ex}.

\vspace{1.0cm}

{\bf \ \ Acknowledgments}

\noindent We thank \'Arp\'ad Heged\H us for providing us with his unpublished
notes. We also thank Sebastien Leurent for a correspondence about problems
related to their NLIE equations.
This work was partially supported by the Hungarian National
Science Fund OTKA (under K116505).

\vspace{1.0cm}

\begin{appendix}

\section{\boldmath Recursion relations for 
$\overline{Q}_0^{(N)}$ and $\overline{Q}_2^{(N)}$}
\label{app_recursion}

In the next subsection we show for the irrep $r$ specified by 
$(m_1,\ldots,m_N)$ and $n_k\equiv m_k+N-k$
$\overline{Q}_0^{(N)}(n_1,\ldots,n_N) = \mathrm{dim}(r)$
and $\overline{Q}_2^{(N)}(n_1,\ldots,n_N)$
satisfy the recursion relations
\begin{equation} \label{Qk_rec}
  \overline{Q}_k^{(N)}(n_1,\ldots,n_N)
  = \sum_{l_1=n_2+1}^{n_1}  \,
  \sum_{l_2=n_3+1}^{n_2} 
  \ldots 
  \sum_{l_{N-1}=n_N+1}^{n_{N-1}}
  \overline{Q}_k^{(N-1)}(l_1,\ldots,l_{N-1})\,,\,\,\,\,k=0,2\,.
\end{equation}
The summation goes over $l_1,\ldots,l_{N-1}$ which satisfy the condition
$n_1 \ge l_1 > n_2 \ge l_2 > \ldots \ge l_{N-2} > n_{N-1} \ge l_{N-1} > n_N$.

To prove the formula for $\overline{C}_2^{(N)}(n_1,\ldots,n_N)$
given by \eqref{CovN} one can then use \eqref{C1Q2}.

\subsection{Proof of the recursion relations}
\label{recursion_proof}

To prove the recursion relations \eqref{Qk_rec} 
we generalize the notion of summation to the case of symbolic limits 
for polynomial summands.

Using the Pochhammer polynomials $(x)_0=1$, $(x)_m=x(x-1)\ldots(x-m+1)$,
for $m=1,2\ldots$ for the finite difference operator 
$\Delta f(x)\equiv f(x+1)-f(x)$ one has
\begin{equation} \label{dif_Pm}
 \Delta\, (x)_m \equiv (x+1)_m - (x)_m = m\, (x)_{m-1} \,'. 
\end{equation}
From this one gets
\begin{equation} \label{sum_Pm}
  \sum_{k=0}^{n-1} (k)_m = \frac{1}{m+1} (n)_{m+1} \,. 
\end{equation}
These are analogous to $\mrd x^m/\mrd x = m x^{m-1}$ and
$\int_0^x t^m \mrd t = x^{m+1}/(m+1)$.

Decomposing a polynomial $P(x)$ by
\begin{equation}
  P(x)=\sum_m c_m\, (x)_m\,, \quad c_m = \frac{1}{m!}\Delta^m P(0)
\end{equation}
one obtains
\begin{equation}
  \sum_{k=0}^{n-1} P(k) = \sum_m \frac{c_m}{(m+1)} (n)_{m+1} \,.
\end{equation}
Define for arbitrary real (or complex) $a$ and $b$
\begin{equation} \label{Sab_def}
  \mathcal{S}(P(x),x,[a,b]) 
  = \sum_m \frac{c_m}{(m+1)} \left[ (b)_{m+1} - (a)_{m+1} \right] \,. 
\end{equation}
For $a,b\in\Z\,,a<b$ one has
\begin{equation}
  \mathcal{S}(P(x),x,[a,b]) = \sum_{k=a}^{b-1} P(k)\,.
\end{equation}
The operation \eqref{Sab_def} can be viewed as an extension of the summation
for polynomials \footnote{This is the rule used by symbolic programs, 
e.g. Maple or Mathematica to evaluate the sum for general limits.}.

With the definition \eqref{Sab_def} one has 
\begin{equation}  \label{Sab_prop}
  \begin{aligned} 
    & \mathcal{S}(P(x),x,[a,a]) = 0 \,, \\
    & \mathcal{S}(P(x),x,[b,a]) = -\mathcal{S}(P(x),x,[a,b]) \,, \\
    & \mathcal{S}(P(x),x,[a,b]) + \mathcal{S}(P(x),x,[b,c])
    = \mathcal{S}(P(x),x,[a,c]) \,, \\
  \end{aligned} 
\end{equation}
which are similar to the properties of $\int_a^b P(x)\mrd x$.

Similarly, we introduce a generalization of multiple sums
\begin{equation}
  \begin{aligned}
    & \mathcal{S}(P(x_1,x_2,\ldots),[x_1,x_2,\ldots] ,
    [[a_1,b_1],[a_2,b_2],\ldots]) 
    = \sum_{k_1=a_1}^{b_1-1} \sum_{k_2=a_2}^{b_2-1}\ldots P(k_1,k_2,\ldots)
    \\
    &\quad = \mathcal{S} \left(\mathcal{S}(P(x_1,x_2,\ldots),[x_2,\ldots] ,
      [[a_2,b_2],\ldots]), x_1, [a_1,b_1]\right) \,.  
  \end{aligned}
\end{equation}
in an analogous way, calculating the rhs for integer values
of the limits with $a_l < b_l$ and using the obtained expression
as a  definition for symbolic limits $a_l$, $b_l$.

Using the invariance under $n_k\to n_k-1$
the recursion relation \eqref{Qk_rec} can for $k=0$ be rewritten as
\begin{equation} 
  \begin{aligned} 
    & \overline{Q}_0^{(N)}(x_1,x_2,x_3,\ldots,x_N)
    \\
    & \qquad
    = \mathcal{S}\left( \overline{Q}Q_0^{(N-1)}(t_1,t_2,\ldots,t_{N-1}),
      [t_1,t_2,\ldots,t_{N-1}], \right. 
    \\
    &\qquad \left. \phantom{\overline{Q}_0^{(N-1)}}
      [[x_1,x_2],[x_2,x_3],[x_3,x_4]\ldots,[x_{N-1},x_{N}]] \right)\,. 
  \end{aligned} 
\end{equation}
Due to \eqref{Sab_prop} this vanishes for $x_1=x_2$ or $x_2=x_3$, etc.

Next, inserting $x_3\to x_1$
\begin{equation} 
  \begin{aligned} 
    & \overline{Q}_0^{(N)}(x_1,x_2,x_1,\ldots,x_N)
    \\
    & \qquad
    = -\mathcal{S}\left(\overline{Q}_0^{(N-1)}(t_1,t_2,\ldots,t_{N-1}),
      [t_1,t_2,\ldots,t_{N-1}],\right. 
    \\
    &\qquad \left. \phantom{\overline{Q}_0^{(N-1)}}
      [[x_1,x_2],[x_1,x_2],\ldots,[x_{N-1},x_{N}]] \right) 
  \end{aligned} 
\end{equation}
Since the limits for $t_1$ and $t_2$ coincide, and the summand
$\overline{Q}_0^{(N-1)}(t_1,t_2,\ldots,t_{N-1})$ changes sign for 
$t_1 \leftrightarrow t_2$, the rhs vanishes for $x_3=x_1$,
and therefore it contains a factor $(x_1-x_3)$.
Obviously, it also contains all factors of type $(x_k-x_{k+2})$.

Consider next the case when $x_4=x_1$.
On the rhs.\ we get the limits
$[[x_1,x_2],$ $[x_2,x_3]$, $[x_1,x_3],\ldots]$.
The range $[x_1,x_3]$ appearing here is the union of
$[x_1,x_2]$ and $[x_2,x_3]$, hence the rhs can be written
as a sum of two expressions with
$[[x_1,x_2], [x_2,x_3], [x_1,x_2],\ldots]$
and 
$[[x_1,x_2], [x_2,x_3], [x_2,x_3],\ldots]$.
Here the ranges $[x_1,x_2]$ (and $[x_2,x_3]$ respectively)
appear twice, and from the antisymmetry wrt. $t_1\leftrightarrow t_2$
(respectively $t_2\leftrightarrow t_3$) one concludes that 
the rhs vanishes also for $x_4=x_1$. 
In this way one can show that the rhs contains all factors $(x_k-x_{k'})$,
$1\le k < k' \le N$ appearing in $\overline{Q}_0^{(N)}(x_1,x_2,x_1,\ldots,x_N)$.
Since the orders of the polynomials on the two sides also coincide,
they must be equal, apart from a possible constant factor,
which can be shown to be 1.
  
\section{The partition function $z_0^{(N)}$ for small $u\,$.}
\label{app_z0_u_small}

Noting
\begin{equation}
  \exp\left[  \frac{u}{2N} \left(\sum_k n_k\right)^2 \right]
  = \sqrt{\frac{N}{2\pi u}}\int_{-\infty}^{\infty} \mrd\alpha
  \exp\left( - \frac{N}{2u}\alpha^2 -\alpha \sum_k n_k   \right)\,,
\end{equation}
we can rewrite \eqref{z0Na0} as
\begin{equation} \label{z0Na2}
  \begin{aligned}
    z_0^{(N)}& = \frac{\mre^{u c_N}}{N!} \sqrt{\frac{N}{2\pi u}}
    \int_{-\infty}^{+\infty} \mrd\alpha \exp\left(-\frac{N}{2u}\alpha^2\right)
    \\
    & \qquad \times
    \left[ \overline{Q}_0^2\left(\frac{\partial}{\partial\alpha_1},
        \ldots,\frac{\partial}{\partial\alpha_{N-1}},0\right)
      \prod_{k=1}^{N-1} \phi(u,\alpha_k)
    \right]_{\alpha_k=\alpha}
  \end{aligned}
\end{equation} 
where
\begin{equation} \label{phiua}
  \phi(u,\alpha) 
  = \sum_{n=-\infty}^\infty 
  \exp\left( -\frac{u}{2} n^2 -\alpha n\right)\,.
\end{equation}
The function $\phi(u,\alpha)$ can be expressed through the Jacobi 
theta-function
\begin{equation} \label{Svz}
  \begin{aligned}
  S(v,z) & = \sum_{n=-\infty}^\infty \mre^{-\pi v (n+z)^2}
  = v^{-1/2} \sum_{n=-\infty}^\infty \mre^{-\pi n^2/v} \cos(2\pi n z) 
  \\
  & = \mre^{-\pi v z^2}  v^{-1/2} S\left(v^{-1}, i v z \right) \,.
  \end{aligned}
\end{equation}
The relation is given by
\begin{equation} \label{phiua1}
  \phi(u,\alpha) = \sqrt{\frac{2\pi}{u}}  
  S\left(\frac{2\pi}{u}, \frac{i\alpha}{2\pi}\right) 
  = \exp\left(\frac{\alpha^2}{2u}\right) 
  S\left( \frac{u}{2\pi}, \frac{\alpha}{u} \right)  \,.
\end{equation}
The function $\phi(u,\alpha)$ satisfies the duality relation 
\begin{equation} \label{phi_phi}
  \phi(u,\alpha)
  = \sqrt{\frac{2\pi}{u}} \exp\left(\frac{\alpha^2}{2u}\right)
  \phi\left( \frac{4\pi^2}{u}, i\frac{2\pi\alpha}{u}\right) \,.
\end{equation}

For $u < 1$ it is convenient to use the fast converging expression
\begin{equation} \label{phiua2}
  \begin{aligned}
    \phi(u,\alpha) 
    &= \sqrt{\frac{2\pi}{u}} \exp\left(\frac{\alpha^2}{2u}\right)
    \sum_{n=-\infty}^\infty \exp\left( -\frac{2\pi^2}{u}n^2\right) 
    \cos\left(\frac{2\pi\alpha}{u} n\right) 
    \\
    &= \sqrt{\frac{2\pi}{u}} \exp\left(\frac{\alpha^2}{2u}\right)
    \left[ 1 + 2 \exp\left( -\frac{2\pi^2}{u}\right) 
    \cos\left(\frac{2\pi\alpha}{u} \right) + \cdots \right]\,. 
  \end{aligned}
\end{equation}
Here the $n\ne 0$ terms in \eqref{phiua2} are 
suppressed exponentially for $u\to 0$. 

%For $u\to 0$ the $n\ne 0$ terms in \eqref{phiua1} are 
%suppressed exponentially.

% Observe that the $N-1$ factors of $\exp(\alpha^2/(2u))$ together with
% $\exp(-N \alpha^2/(2u))$ still give a factor $\exp(-\alpha^2/(2u))$
% hence the integral over $\alpha$ in \eqref{z0Na} is well defined.
% Inserting the expansion \eqref{phiua2} the resulting integrals
% in $\alpha$ are Gaussian, and one obtains an expansion
% in powers of $\exp(-2\pi^2/u)$.

For $u\ll1$ and defining $w=4\pi^2/u\,,$ one obtains for $N=2,3,4\,,$
the expansions 
\begin{equation} \label{z0N}
  \begin{aligned}
    z_0^{(2)}(u) & = 2\sqrt{\pi} \mre^{u/4} u^{-3/2}
    \left[ 1 - 2\,\mre^{-w} (2w - 1) + \ldots
    \right] \,, \\
    z_0^{(3)}(u) & = \sqrt{3} \pi \mre^{u} u^{-4} 
    \left[1 - \mre^{-w} (2 w^3-9 w^2
      +18 w -6) + \ldots \right] \,, \\
    z_0^{(4)}(u) & = \frac{\sqrt{2}}{3} \pi^{3/2} \mre^{5u/2} u^{-15/2} 
    \left[1 - \frac16 \,\mre^{-w} \left( 2 w^5
      -25 w^{4} \right. \right. \\
      & \quad \left. \phantom{\frac16} \left. +128 w^{3}
        -276 w^{2}+288 w -72\right) + \ldots \right] 
    % \,, \\
    % z_0^{(5)}(u) & = \frac{\sqrt{5}}{72} \pi^{2} \mre^{5u} u^{-12} 
    % \left[ 1 + \order{\mre^{-w} u^{-7}}\right]
    \,.
  \end{aligned}
\end{equation}

\section{The partition function $Z(u;h)$}
\label{app_Zuh}

In this appendix we consider the full dependence of the partition function
on the chemical potential $h$. 
Below we use the short-hand notation $\hh=h L_t$

\subsection{U(1) case}

The irreps of the U(1) group are labeled by $m\in \mathbb{Z}$, and
all have dimension 1, while the Casimir invariant is $C=m^2\,$.
The partition function is 
\begin{equation}
  Z(u;h)= \sum_{m=-\infty}^\infty \mre^{-u m^2 -m \hh}
  = \exp\left(\frac{\hh^2}{4u}\right) 
  S\left( \frac{u}{\pi}, \frac{\hh}{2u} \right)\,.
\end{equation}

\subsection{SU(2) case}

For SU(2) the partition function is given by
\begin{equation}
  \begin{aligned}
    Z(u;h) & = \sum_{p=0}^{\infty}\,\,
    \exp\left(-\frac14 p(p+2) u\right)
    \left[\sum_{m=-p/2}^{p/2} \mre^{-m \hh }  \right]^2
    \\
    % & = \sum_{p=0}^{\infty} \exp\left(-\frac14 p(p+2) u\right)
    % \left(\frac{\sinh(\hh (p+1)/2)}{\sinh(\hh /2)}\right)^2
    % \\
    & = \mre^{u/4}\sum_{n=1}^{\infty} \exp\left(-\frac14 n^2 u\right)
    \left(\frac{\sinh(\hh n/2)}{\sinh(\hh /2)}\right)^2\,.
  \end{aligned}
\end{equation}
Simplifying one gets
\begin{equation} \label{z0uh_N2}
  Z(u;h) = \frac{\mre^{u/4}}{4 \sinh^2(\hh /2)}
  \left[ \mre^{\hh^2/u} S\left(\frac{u}{4\pi},\frac{2\hh }{u}\right)
    - S\left(\frac{u}{4\pi},0\right) \right]\,.
\end{equation}
In the limit $h\to 0:\,\,$  $z_0^{(N)}(u) = \lim_{h\to 0} Z(u;h)$ 
thereby recovering \eqref{z0_SU2}.

\subsection{$\SUN$ case}

For general $N$ we have
\begin{equation}
  Z(u;h) = \sum_{r}\,\, \exp\left(-u C_2^{(N)}(r)\right)
  \left[R(r;\hh) \right]^2\,,
\end{equation}
where the sum goes over irreps $r$, and 
\begin{equation}
  R(r;\hh) = \mathrm{Tr}_r \exp\left(-\hh  J_3\right)
  =\sum_{s\in r}\,\, \exp\left(-\lambda_1(s) \hh \right)
\end{equation}
where $s$ runs over an appropriate basis of the representation $r$ of 
$\SUN$, and $\lambda_1(s)$ is the corresponding eigenvalue of $J_3$.

Using the convention $n_k = m_k + N-k$ one can write  $R(r;\hh)$
in the form
\begin{equation} \label{Rbar}
  \overline{R}(n_1,\ldots,n_N) =
  \rho(\hh) \sum_l
    b^{(l)} \sinh\left( \hh  \sum_{k=1}^N a_k^{(l)} n_k \right)\,.
\end{equation}
The coefficients $a_1^{(l)},\ldots,a_N^{(l)}$, $b^{(l)}$
and $\rho(\hh)$ 
will be calculated below explicitly for $\SUth$.

From this one obtains
\begin{equation} \label{Rbarsq}
  \begin{aligned}
    \left[\overline{R}(n_1,\ldots,n_N)\right]^2 &=
    \frac12 \rho^2(\hh) \sum_{l l'} b^{(l)}b^{(l')}
    \\
    &
    \left[ 
      \cosh\left( \sum_{k=1}^N (a_k^{(l)}+a_k^{(l')})\hh  n_k \right) 
      -\cosh\left( \sum_{k=1}^N (a_k^{(l)}-a_k^{(l')})\hh  n_k \right) 
    \right]\,.
  \end{aligned}
\end{equation}

Similarly to the steps in \eqref{z0Na2} one obtains
\begin{equation} \label{z0uhN}
  Z(u;h) = \frac{\mre^{u c_N}}{N!}\sqrt{\frac{N}{8\pi u}}\,
  \rho^2(\hh) 
  \int_{-\infty}^{+\infty} \mrd\alpha \exp\left(-\frac{N}{2u}\alpha^2\right)
  \Psi(u,\alpha,\hh)
\end{equation}
where
\begin{multline} \label{Psiuah}
    \Psi(u,\alpha,\hh) =
    \sum_{l l'} b^{(l)}b^{(l')}
    \left\{ 
      \prod_{k=1}^{N-1} \phi\left(u,\alpha+\hh (a_k^{(l)}+a_k^{(l')})\right)
      \right.
      \\
      \left.
      -\prod_{k=1}^{N-1} \phi\left(u,\alpha+\hh (a_k^{(l)}-a_k^{(l')})\right)
    \right\}
\end{multline}
where $\phi(u,\alpha)$ is as in \eqref{phiua1}\,. 
For small $u$ one can use the expansion \eqref{phiua2}.

\subsection{SU(2) case, again}

Using $m_1\equiv m_{12}$, $m_2\equiv m_{22}$ 
\begin{equation}
  R(m_1,m_2;\hh) = 
  \sum_{m_{11}=m_2}^{m_1}
    \exp\left\{-\hh  \left( m_{11}-\frac12 (m_{1}+m_{2})\right)\right\}\,. 
\end{equation}
With $n_1=m_1+1$, $n_2=m_2$ and $n_{11}=m_{11}$ this gives
\begin{equation}
  \overline{R}(n_1,n_2;\hh) = 
  \sum_{n_{11}=n_2}^{n_1-1}
    \exp\left\{-\hh  \left( n_{11}-\frac12 (n_1+n_2-1)\right)\right\}
    =\frac{\sinh((n_1-n_2)\hh /2)}{\sinh(\hh /2)}\,.
\end{equation}
Hence $\rho(\hh)=1/\sinh(\hh /2)$ and we have only one term, $l=1$ 
with $b^{(1)}=1$ and $a^{(1)} = \left[\frac12, -\frac12\right]\,.$

From \eqref{z0uhN} we get
\begin{equation}
 Z(u;h) = \frac{\mre^{u/4}}{4 \sinh^2\frac{\hh }{2}}\sqrt{\frac{1}{\pi u}}
 \int_{-\infty}^{\infty}\mrd\alpha \mre^{-\alpha^2/u}
 \left[ \phi(u,\alpha+\hh )- \phi(u,\alpha)\right]\,.
\end{equation}

Using for $\phi(u,\alpha)$ the expansion \eqref{phiua} one gets
\begin{equation}
\begin{aligned}
  Z(u,h) &= \frac{\mre^{u/4}}{4\sinh^2\frac{\hh }{2}}
  \sqrt{\frac{1}{\pi u}}
  \int_{-\infty}^{\infty}\mrd\alpha\mre^{-\alpha^2/u}
  \sum_{n=-\infty}^\infty \mre^{-u n^2/2-\alpha n} 
\left(\mre^{-\hh  n}-1\right)
\\
   &= \frac{\mre^{u/4}}{4\sinh^2\frac{\hh }{2}}
  \sum_{n=-\infty}^\infty \mre^{-u n^2/4} \left(\mre^{-\hh  n}-1\right)\,.
\end{aligned}
\end{equation}
Further
\begin{equation}
  \sum_n \mre^{-u n^2/4} \mre^{-\hh  n} = 
  \mre^{\hh^2/u} \sum_n \mre^{-u (n+2\hh /u)^2/4}
  = \mre^{\hh^2/u} S\left(\frac{u}{4\pi},\frac{2\hh }{u}\right)\,,
\end{equation}
and inserting this one recovers \eqref{z0uh_N2}.

\subsection{SU(3) case}
\label{Zuh_SU3}

With the notation $m_k\equiv m_{k3}$ one has
\begin{equation}
  R(m_1,m_2,m_3;\hh) = 
  \sum_{m_{12}=m_2}^{m_1} \sum_{m_{22}=m_3}^{m_2} 
  \sum_{m_{11}=m_{22}}^{m_{12}} 
    \exp\left\{-\hh  \left( m_{11}-\frac12 (m_{12}+m_{22})\right)\right\}\,. 
\end{equation}
With $n_{kM}=m_{kM}+M-k$ and $n_{kN}\equiv n_k$ this gives
\begin{equation}
  \begin{aligned}
    & \overline{R}(n_{1},n_{2},n_{3};\hh)
    = \sum_{n_{12}=n_{2}}^{n_{1}-1}\,\,
    \sum_{n_{22}=n_{3}}^{n_{2}-1}\,\,\sum_{n_{11}=n_{22}}^{n_{12}-1}\,\,
    \exp\left\{-\hh  \left( n_{11}-\frac12 (n_{12}+n_{22}-1)\right)
    \right\}
    \\
    & \quad = \frac{\sinh\left((n_{2}-n_{1})\hh /2\right)
      +\sinh\left((n_{3}-n_{2})\hh /2\right)
      +\sinh\left((n_{1}-n_{3})\hh /2\right)}%
    {2\sinh\frac{\hh }{2} \left(\cosh\frac{\hh }{2}-1\right)}
  \end{aligned}
\end{equation}
As a check one has
\begin{equation}
  \lim_{\hh\to 0}\overline{R}(n_{1},n_{2},n_{3};\hh) =
  \frac12 (n_{1}-n_{2}) (n_{1}-n_{3})(n_{2}-n_{3})
  = \overline{Q}_0(n_{1},n_{2},n_{3})\,. 
\end{equation}
One gets
\begin{equation}
  \rho(\hh) = \frac{1}{4\sinh\frac{\hh }{2} \sinh^2\frac{\hh }{4}}
  = 8 \hh^{-3} + \order{\hh^{-4}}\,.
\end{equation}
We have 3 terms, $l=1,2,3$, with $b^{(l)}=1$ and the coefficients 
$[a^{(1)}_1,a^{(1)}_2,a^{(1)}_3]$ are
\begin{equation} \label{aaa}
  a^{(1)} = \left[-\frac12, \frac12,0\right] \,, \quad
  a^{(2)} = \left[0, -\frac12, \frac12 \right] \,, \quad
  a^{(3)} = \left[\frac12, 0, -\frac12\right] \,.
\end{equation}

One has (using the symmetry $h\to -h$)
\begin{equation}  
  \begin{aligned}
    \Psi(u,\alpha,\hh) &= 
    \phi(u,\alpha-\hh )\phi(u,\alpha+\hh )
    +2\phi(u,\alpha)\phi(u,\alpha+\hh )
    \\
    & \quad
    +4\phi(u,\alpha)\phi(u,\alpha+\hh /2)
    +\phi(u,\alpha+\hh /2)\phi(u,\alpha-\hh /2)
    \\
    & \quad
    -4\phi(u,\alpha+\hh )\phi(u,\alpha-\hh /2)
    -\phi(u,\alpha+\hh /2)^2 
    -3 \phi(u,\alpha)^2\,.
  \end{aligned}
\end{equation}
Neglecting exponentially small terms $\order{\exp(-2\pi^2/u)}$ 
in \eqref{phiua2} one obtains
\begin{equation}  
  \begin{aligned}
    Z(u;h) & \simeq \sqrt{3} \pi \mre^{u}
    \frac{\left(\exp(\frac{\hh^2}{2 u})-1\right)
      \left(\exp(\frac{\hh^2}{4 u})-1\right)^2}{%
      32 u \sinh^2\frac{\hh }{2}\sinh^4\frac{\hh }{4}}
    \\
    & = \sqrt{3} \pi \mre^{u} u^{-4}
    \left[ 1 + \frac12 \hh^2 \left(\frac{1}{u}-\frac14 \right)
      + \order{\hh^4} \right]\,.
  \end{aligned}
\end{equation}
This is in agreement with \eqref{Zuh}, \eqref{z0N}, \eqref{z1z0_SUN}.

\section{Relation of $z_0(u)$ to the heat kernel $K(U,t)$}

The heat kernel $K(U,t)$ on the group manifold $\SUN$ in the $U\to I$
limit (where $I$ is the identity) is related 
to our partition function $z_0(u)$ at $u=t$.
It is given by (see eqs.~(3), (6) from \cite{Menotti:1981ry})
\begin{equation} \label{KUt}
  K(U,t) = \langle I | \mre^{-t \hat{J}^2 } | U \rangle =
  \sum_r d^{(r)} \chi^{(r)}(U) \exp\left(-t \mathcal{C}_2^{(r)}\right)\,,
\end{equation}
where $r$ runs over all irreducible unitary representations,
$\chi^{(r)}(U)$ is the character of the representation,
$d^{(r)}= \chi^{(r)}(I)$ its dimension and $\mathcal{C}_2^{(r)}$
its quadratic Casimir invariant. 

Formally one has
\begin{equation}
  K(I,t)
  = \langle I | \exp(-t \hat{L}^2) | I \rangle 
  = \int_U \langle U | \exp(-t \hat{L}^2) | U \rangle 
  = \mathrm{Tr}( \exp(-t\hat{L}^2 ) )
  = z_0(t)\,.
\end{equation}
From \eqref{KUt} one also has
\begin{equation}
  K(I,t) = \sum_r [d^{(r)}]^2\exp\left(-t \mathcal{C}_2^{(r)}\right)
  = z_0(t)\,.
\end{equation}

For SU(2), eq.~(9),\cite{Menotti:1981ry}
\begin{equation}
  K(\phi,t) = \mathcal{N}_2 \sum_{n=-\infty}^\infty
  \frac{\phi + 2\pi n}{\sin{\phi}} 
  \exp\left(-\frac{(\phi + 2\pi n)^2}{t}\right)
\end{equation}
where $U=S \, \mathrm{diag}[\phi,-\phi]\, S^\dagger$, and
the prefactor $\mathcal{N}_2$ does not depend on $\phi$.
Note, however, that it depends on the parameter $t$, a fact 
which was irrelevant for the discussion of \cite{Menotti:1981ry}).

In the limit $U\to I$ this gives
\begin{equation}
  \lim_{\phi\to 0} K(\phi,t) = \mathcal{N}_2(t) 
  \left[ S\left(\frac{4\pi}{t}\right)
    + \frac{8\pi}{t}S'\left(\frac{4\pi}{t} \right)
  \right]\,.
\end{equation}
From our result in eq.~\eqref{z0_SU2_A} we deduce
\begin{equation}
  \mathcal{N}_2(t) = \sqrt{4\pi} \mre^{t/4} t^{-3/2} \,.
\end{equation}
Note that for $t\to 0$ the square bracket goes to $1$ exponentially
fast and the important information for the isospin susceptibility
in this limit is hidden entirely in the undetermined
$t$-dependence of $\mathcal{N}(t)$.

For $\SUN$ one has from eq.~(7),\cite{Menotti:1981ry}
\begin{equation}
  z_0(t)=\lim_{\phi\to 0} K(\phi,t) = \mathcal{N}_N(t)
  \sum_{\{l\}=-\infty}^{\infty}
  \prod_{i<j}\left[ 1 - \frac{4\pi^2}{t N}(l_i-l_j)^2\right]
  \exp\left( - \frac{2\pi^2}{t N} (l_i-l_j)^2\right)\,.
\end{equation}
Comparing this with \eqref{z0NN} one gets
\begin{equation}
  \mathcal{N}_N(t) 
  = \frac{(2\pi)^{(N-1)/2}\sqrt{N}}{2!\,3!\ldots (N-1)!}
    t^{-(N^2-1)/2} \exp\left( c_N t\right)\,. 
\end{equation}

%\input TY.tex

%%%%%%%%%%%%%%%%%%%%%%%%%%%%%%%%%%%%%%%%%%%%%%%%%%%%%%%%%%%%%%%%%%%%%%%%%%%%%%
%%%%%%%%%%%%%%%%%%%%%%%%%%%%%%%%%%%%%%%%%%%%%%%%%%%%
%%%%%%%%%%      SU3 Y-system     %%%%%%%%%%%%%%%%%

\newsavebox{\SSa}
\sbox{\SSa}{
\setlength{\unitlength}{1.2mm}
\begin{picture}(140,25) (-30,-12.5)

\put(0,0){\circle*{3}}
\put(10,0){\circle{3}}
\put(20,0){\circle{3}}
\put(30,0){\circle{3}}
\put(40,0){\circle{3}}
\put(-10,0){\circle{3}}
\put(-20,0){\circle{3}}
\put(-30,0){\circle{3}}
\put(-40,0){\circle{3}}

\put(0,7){\circle*{3}}
\put(10,7){\circle{3}}
\put(20,7){\circle{3}}
\put(30,7){\circle{3}}
\put(40,7){\circle{3}}
\put(-10,7){\circle{3}}
\put(-20,7){\circle{3}}
\put(-30,7){\circle{3}}
\put(-40,7){\circle{3}}

\put(1.5,0){\line(1,0){7}}
\put(11.5,0){\line(1,0){7}}
\put(21.5,0){\line(1,0){7}}
\put(31.5,0){\line(1,0){7}}
\put(41.5,0){\line(1,0){4}}
\put(-8.5,0){\line(1,0){7}}
\put(-18.5,0){\line(1,0){7}}
\put(-28.5,0){\line(1,0){7}}
\put(-38.5,0){\line(1,0){7}}
\put(-45.5,0){\line(1,0){4}}

\put(1.5,7){\line(1,0){7}}
\put(11.5,7){\line(1,0){7}}
\put(21.5,7){\line(1,0){7}}
\put(31.5,7){\line(1,0){7}}
\put(41.5,7){\line(1,0){4}}
\put(-8.5,7){\line(1,0){7}}
\put(-18.5,7){\line(1,0){7}}
\put(-28.5,7){\line(1,0){7}}
\put(-38.5,7){\line(1,0){7}}
\put(-45.5,7){\line(1,0){4}}

\put(0,1.5){\line(0,1){4}}
\put(10,1.5){\line(0,1){4}}
\put(20,1.5){\line(0,1){4}}
\put(30,1.5){\line(0,1){4}}
\put(40,1.5){\line(0,1){4}}
\put(-10,1.5){\line(0,1){4}}
\put(-20,1.5){\line(0,1){4}}
\put(-30,1.5){\line(0,1){4}}
\put(-40,1.5){\line(0,1){4}}

%\put(31.1,1.1){\line(1,1){7.7}}
%\put(31.1,-1.1){\line(1,-1){7.7}}
%\put(-31.1,1.1){\line(-1,1){7.7}}
%\put(-31.1,-1.1){\line(-1,-1){7.7}}

\multiput(44.1,3.5) (1,0) {4} {\circle*{0.2}}
\multiput(-48.5,3.5) (1,0) {4} {\circle*{0.2}}

\put(0,-3){\makebox(0,0)[t]{{\protect\scriptsize 0}}}
\put(10,-3){\makebox(0,0)[t]{{\protect\scriptsize 1}}}
\put(20,-3){\makebox(0,0)[t]{{\protect\scriptsize 2}}}
\put(30,-3){\makebox(0,0)[t]{{\protect\scriptsize 3}}}
%\put(22,-3){\makebox(0,0)[t]{{\protect\scriptsize {\em p}--2}}}
%\put(30,-3){\makebox(0,0)[t]{{\protect\scriptsize {\em p}--1}}}
%\put(44,-11){\makebox(0,0)[t]{{\protect\scriptsize {\em p}}}}
%\put(47,11){\makebox(0,0)[t]{{\protect\scriptsize {\em p}+1}}}
\put(-10,-3){\makebox(0,0)[t]{{\protect\scriptsize --1}}}
\put(-20,-3){\makebox(0,0)[t]{{\protect\scriptsize --2}}}
\put(-30,-3){\makebox(0,0)[t]{{\protect\scriptsize --3}}}
%\put(-18,-3){\makebox(0,0)[t]{{\protect\scriptsize 
%\underline{{$\tilde{p}$ }--2}}}}
%\put(-28,-3){\makebox(0,0)[t]{{\protect\scriptsize 
%\underline{{$\tilde{p}$ }--1}}}}
%\put(-45,-10){\makebox(0,0)[t]{{\protect\scriptsize 
%\underline{{$\tilde{p}$}}}}}
%\put(-46,11){\makebox(0,0)[t]{{\protect\scriptsize 
%\underline{{$\tilde{p}$}+1}}}}

\end{picture}}

%%%%%%%%%%%%%%%%%%%%%%%%%%%%%%%%%%%%%%%%%%%%%%%%%%%%%%
%%%%%%%%%%%%%%%%%%%%%%%%%%%%%%%%%%%%%%%%%%%%%%%%%%%%%%%%%%%%%%%%%%%%%%%%%%%

\section{Hirota dynamics and NLIE for the SU$(3)$ principal model}

In this appendix we give formulas necessary to numerically compute the finite
volume energy spectrum of the SU$(N)$ principal chiral model for $N=3$ using
the NLIE equations constructed in Ref. \cite{Kazakov}. \cite{Kazakov} 
discusses the case
for general $N\geq3$, but here for simplicity we restrict our attention 
to $N=3$ only. 
(The $N=2$ case was discussed before in \cite{KL0}.) We give all formulas
necessary to perform the numerical computation in a \lq\lq cookbook style'' and
refer to the original paper \cite{Kazakov} for the derivation of the 
equations and further details.

\subsection{SU$(3)$ T-system and Y-system}

Based on previous experience with integrable models, where similar equations
were constructed by starting from integrable lattice regularizations and/or
by bootstrap methods the following double-infinite T-system is proposed as the
basis for the description of the finite volume spectrum of the SU$(3)$ principal
model:
\begin{equation}
T^+_{a,s}(\theta)\,T^-_{a,s}(\theta)=
T_{a,s+1}(\theta)\,T_{a,s-1}(\theta)+
T_{a+1,s}(\theta)\,T_{a-1,s}(\theta),
\label{Tsys}
\end{equation}
where the T-functions $T_{a,s}(\theta)$ are indexed by $a=0,1,2,3$ and
$s=0,\pm1,\pm2,\dots$ and by definition
\begin{equation}
T_{-1,s}(\theta)=T_{4,s}(\theta)\equiv0\,.
\end{equation}
Here for any function $f(\theta)$ the notation $f^\pm(\theta)$ stands for
\begin{equation}
f^\pm(\theta)=f\left(\theta\pm\frac{i}{2}\right)\,.  
\end{equation}
For the description of a particular state in the spectrum of the model we have
to specify the corresponding solution of the T-system (\ref{Tsys}). Starting
from the T-system (Hirota equations) one can go to the corresponding
double-infinite Y-system, which is used to construct the TBA integral
equations, or, alternatively, to the finite Q-system, which is used in the
NLIE approach.

The SU$(3)$ Y-system for the Y-functions $Y_{a,s}(\theta)\,$, $a=1,2$,
$s=0,\pm1,\pm2,\dots$ is
\begin{equation}
Y^+_{a,s}(\theta)\,Y^-_{a,s}(\theta)=
[1+Y_{a,s+1}(\theta)][1+Y_{a,s-1}(\theta)]\,
\frac{Y_{a+1,s}(\theta)}{1+Y_{a+1,s}(\theta)}\,
\frac{Y_{a-1,s}(\theta)}{1+Y_{a-1,s}(\theta)},
\label{Ysys}
\end{equation}
with the convention
\begin{equation}
Y_{0,s}(\theta)=Y_{3,s}(\theta)\equiv\infty\,.
\end{equation}

%%%%%%%%%%%%%%%%%%%%%%%%%%%%%%%%%%%%%%%%%%%%%%%%%%%%%%%%%%%%%%%%%%%%%%%%%%%
%\vspace{1cm}
\begin{figure}[tbp]
\label{SU3TBA}
\begin{center}
\begin{picture}(140,30)(0,-15)
\put(-35,-50) {\usebox{\SSa}}
%\put(-1,-7){\parbox{130mm}{\caption{ \label{TBA}\protect {\small
%TBA-diagram associated with the SU(3)-model Y-system. }}}}
\end{picture}
\end{center}
\caption{\footnotesize TBA-diagram associated with the SU(3) model Y-system.}
\end{figure}
%%%%%%%%%%%%%%%%%%%%%%%%%%%%%%%%%%%%%%%%%%%%%%%%%%%%%%%%%%%%%%%%%%%%%%%%%%%%%%%

This Y-system is illustrated by Fig. 2, where the $s=0$ nodes are
black indicating that they are the massive nodes with asymptotic behavior
\begin{equation}
Y_{a,0}(\theta)\sim{\rm e}^{-ML\cosh(v\theta)}\cdot{\rm const.}
\qquad\quad \vert\theta\vert\longrightarrow\infty,    
\end{equation}
where
\begin{equation}
v=\frac{2\pi}{3}\,.
\end{equation}
Here $M$ is the mass of the particles in infinite volume and $L$ is the size
of the system.
All other (magnonic) nodes behave as
\begin{equation}
Y_{a,s}(\theta)\sim{\rm const.}\qquad\quad
\vert\theta\vert\longrightarrow\infty,\qquad s\not=0\,.    
\end{equation}

The relation between the T-functions and Y-functions is
\begin{equation}
Y_{a,s}(\theta)=\frac{T_{a,s+1}(\theta)\,T_{a,s-1}(\theta)}
{T_{a+1,s}(\theta)\,T_{a-1,s}(\theta)}\,,\qquad\qquad
1+Y_{a,s}(\theta)=\frac{
T^+_{a,s}(\theta)\,T^-_{a,s}(\theta)}
{T_{a+1,s}(\theta)\,T_{a-1,s}(\theta)}\,.
\label{TY}
\end{equation}

We will consider the U$(1)$ sector only, where all particles are highest weight
states in the defining representation of SU$(3)$. For these states one can
establish, using the T-Y relations (\ref{TY}) that
\begin{equation}
\begin{split}  
1+Y_{1,0}(\theta) \ {\rm has\  zeroes\  at:}\ \theta&=\theta_{1,j}+\frac{3i}{4};
\quad \theta=\theta_{1,j}-\frac{i}{4},\\
1+Y_{1,0}(\theta) \ {\rm has\  poles\  at:}\ \theta&=\theta_{2,j}-\frac{i}{4}\\
\end{split}
\label{Y10poles}
\end{equation}
and
\begin{equation}
\begin{split}  
1+Y_{2,0}(\theta) \ {\rm has\  zeroes\  at:}\ \theta&=\theta_{2,j}-\frac{3i}{4};
\quad \theta=\theta_{2,j}+\frac{i}{4},\\
1+Y_{2,0}(\theta) \ {\rm has\  poles\  at:}\ \theta&=\theta_{1,j}+\frac{i}{4}.\\
\end{split}
\label{Y20poles}
\end{equation}
The position of singularities are not independent; they are related by the
T-Y relations. They are parameterized in terms of two complex quantities
$\theta_{1,j}$ and $\theta_{2,j}$, which are deformations of (and for large $L$
are exponentially close to) the real asymptotic rapidities (Bethe roots)
$\theta_j$. The index $j$ ($j=1,\dots,{\cal N}$) labels the particles.
The two Y-functions are conjugates of each other:
\begin{equation}
\left[Y_{1,s}(\theta)\right]^*=Y_{2,s}(\theta^*)
\label{conj12}
\end{equation}
and consequently
\begin{equation}
\theta_{1,j}^*=\theta_{2,j}.  
\end{equation}

Whether one uses the infinite set of TBA equations, which can be derived from
the Y-system (\ref{Ysys}), or the finite set of NLIE equations of
Ref. \cite{Kazakov}, it is always necessary to construct at least the $s=0$
Y-functions corresponding to the massive nodes, since these are entering the
energy formula\footnote{Note that in this appendix the symbol $E$ generically
denotes the energy of the given state and not as in the main text the
energy gap between this energy level and the ground state energy.}  
\begin{equation}
\begin{split}
E=m&\sum_{j=1}^{\cal N}\left\{
\cosh\left[v\left(\theta_{1,j}+\frac{i}{2}\right)\right]+
\cosh\left[v\left(\theta_{2,j}-\frac{i}{2}\right)\right]\right\}\\
-\frac{m}{3}&\int_{-\infty}^\infty{\rm d}\theta\cosh(v\theta)\,\ln\left(
[1+Y_{1,0}(\theta)][1+Y_{2,0}(\theta)]\right)\,.
\label{ener}
\end{split}
\end{equation}  
This energy formula was conjectured in Ref. \cite{Kazakov}, but it can also be
systematically derived~\cite{Heg} from an integrable lattice regularization
of the model.

\subsection{Asymptotic solutions}

We are not going to use the infinite set of TBA equations in our numerical
calculations but we used them to derive some large volume asymptotic formulas
to determine the exponentially small shifts of the parameters $\theta_{a,j}$
and to calculate the first exponentially small corrections to the energy
(\ref{ener}). In this subsection we will use the \lq\lq natural'' normalization
of the rapidity parameters (to avoid confusion we denote them by $T$ instead
of $\theta$). The relation between the two normalizations is
\begin{equation}
T_{a,j}=v\theta_{a,j},\qquad\quad T_j=v\theta_j.
\end{equation}

To leading order the energy is just the sum of the individual free particle
energies given by
\begin{equation}
E^{(0)}=\sum_{j=1}^{\cal N}M\cosh T_j\,,  
\end{equation}
where the asymptotic rapidities satisfy the Bethe quantization conditions
\begin{equation}
2\pi n_j=ML\sinh T_j+\sum_{i\not=j}\delta(T_j-T_i)\,,  
\end{equation}
where $n_j$ are integer quantum numbers and $\delta(\theta)$ is the phase shift
for the scattering of highest weight triplet particles. In the simplest
1-particle case the quantization condition reduces to
\begin{equation}
\sinh T_1=\frac{2\pi n_1}{ML}\,.  
\end{equation}

The next (leading exponential, also called L\"uscher) corrections consist of two
parts:
\begin{equation}
E^{({\rm L})}=E^{(\mu)}+E^{(F)}\,,
\end{equation}
where the mu-term $E^{(\mu)}$ comes from the deformation of the rapidity
parameters in the first line of (\ref{ener}) and the F-term $E^{(F)}$ is
the leading exponential approximation of the second line (the integral term).
We can write the change of the rapidities as  
\begin{equation}
T_{1,j}=T_j+x_j-iy_j,\qquad\quad T_{2,j}=T_j+x_j+iy_j\,,  
\end{equation}
where both $x_j$ and $y_j$ are exponentially small. The sign of $y_j$ determines
whether when moving away from the $L=\infty$ limit the exact rapidities move
upwards or downwards in the complex plane.

We now write down the L\"uscher order asymptotic formulas for ${\cal N}=1$ and
for the lowest energy zero total momentum ${\cal N}=2$ state.

First we introduce some notations and definitions.
\begin{equation}
\lambda(\theta)={\rm e}^{2ib(\theta)},\qquad\quad b(\theta)=\frac{\pi}{2}-
\arctan\left(\frac{1}{\sqrt{3}}\tanh\frac{\theta}{2}\right),
\end{equation}
\begin{equation}
a(\theta)=\sqrt{\frac{3}{4}+\sinh^2\left(\frac{\theta}{2}\right)},\qquad
\Gamma\left(1+\frac{i\theta}{2\pi}\right)\,  
\Gamma\left(\frac{1}{3}-\frac{i\theta}{2\pi}\right)=  
A(\theta){\rm e}^{iB(\theta)}\,,
\label{AB}
\end{equation}
\begin{equation}
{\cal L}(\theta)=\frac{{\rm e}^{2iB(\theta)}}{2+\frac{3i\theta}{\pi}},\qquad  
D(\theta)=\left\vert\Gamma\left(\frac{2}{3}+\frac{i\theta}{2\pi}\right)
\right\vert^2\,,  
\end{equation}
\begin{equation}
{\cal K}(\theta)=\frac{a(\theta)\,D^2(\theta)}
{\sinh\frac{\theta}{2}\,A^2(\theta)},\qquad  
{\cal A}=\frac{64\pi^4}{9\Gamma^6(1/3)}\,.
\end{equation}
We note that in (\ref{AB}) $A(\theta)$ and $B(\theta)$ are real for $\theta$
real.

Further
\begin{equation}
\mu(\theta)=\lambda(\theta)\left[{\cal L}(\theta)
\left(2+\frac{9i\theta}{\pi}\right)\right]^2\,,
\end{equation}
\begin{equation}
g(\alpha,\beta)=\lambda(\alpha)\lambda(\beta)\left(
{\cal L}(\alpha){\cal L}(\beta)\left[
4+\frac{6i}{\pi}(\alpha+\beta)-\frac{27}{\pi^2}\alpha\beta\right]\right)^2\,,
\end{equation}
\begin{equation}
f(\alpha,\beta)=g\left(\alpha+\frac{i\pi}{2},\beta+\frac{i\pi}{2}\right),
\qquad\quad f_1(\alpha,\beta)=\frac{\partial}{\partial\alpha}
f(\alpha,\beta)\,.  
\end{equation}

For ${\cal N}=1$ we find
\begin{equation}
y_1=-\sigma{\cal A}(-1)^{n_1}\,{\rm e}^{-\sigma z\cosh T_1}\,,
\label{N1y1}
\end{equation}
\begin{equation}
E^{(\mu)}=-\frac{32M\pi^4(-1)^{n_1}}{3\Gamma^6(1/3)\cosh T_1}\,
{\rm e}^{-\sigma z\cosh T_1}\,,
\label{N1mu}
\end{equation}
\begin{equation}
E^{(F)}=-\frac{M}{2\pi\cosh T_1}\int_{-\infty}^\infty{\rm d}\theta\, \cosh\theta
\,{\rm e}^{-z\cosh(\theta+T_1)}\left[\mu\left(\theta+\frac{i\pi}{2}\right)+
\mu\left(\frac{i\pi}{2}-\theta\right)\right]\,.
\end{equation}
Here
\begin{equation}
\sigma=\frac{\sqrt{3}}{2},\qquad\quad z=ML\,.
\end{equation}
The coefficient of the exponential in the mu-term (\ref{N1mu}) (for the
standing particle $T_1=0$) is different from that given by Eq.~(95) of
Ref.~\cite{Kazakov}.
Although our coefficient is larger by a factor $\pi/3\approx 1.05$ only, we
have demonstrated numerically that the difference between the two formulas is
clearly visible and that it is indeed (\ref{N1mu}) that agrees asymptotically
with the exact result.

For the lowest energy parity symmetric ${\cal N}=2$ state with
\begin{equation}
T_1=\bar T=-T_2>0
\end{equation}
we find
\begin{equation}
y_1=y_2=-\sigma {\cal A}\,{\rm e}^{-\sigma z\cosh \bar T}\,{\cal K}(2\bar T)\,,
\label{N2y12}
\end{equation}
\begin{equation}
E^{(\mu)}=-\frac{M{\cal A}}{\cosh\bar T}\,{\rm e}^{-\sigma z\cosh\bar T}
\left\{3{\cal K}(2\bar T)+\frac{8\sigma}{z}\sinh\bar T\,{\cal K}^\prime(2\bar T)
\right\}\,,
\end{equation}
\begin{equation}
\begin{split}
E^{(F)}=-\frac{M}{\pi}\int_{-\infty}^\infty{\rm d}\theta\,{\rm e}^{-z\cosh\theta}
&\Big\{\cosh\theta f(\theta-\bar T,\theta+\bar T)\\  
+\frac{\tanh \bar T}{z}&\left[f_1(\theta+\bar T,\theta-\bar T)
-f_1(\theta-\bar T,\theta+\bar T)\right]\Big\}\,.
\end{split}
\end{equation}

\subsection{Alternative energy formula}

In Ref. \cite{Kazakov} an alternative energy formula is used:
\begin{equation}
\begin{split}
E_{\rm KL}=-\frac{M}{3}\int_{-\infty}^\infty{\rm d}\theta\Big\{
&\cosh v\left(\theta-\frac{i}{4}\right)\ln\left[1+Y_{1,0}
\left(\theta-\frac{i}{4}\right)\right]\\  
+&\cosh v\left(\theta+\frac{i}{4}\right)\ln\left[1+Y_{2,0}
\left(\theta+\frac{i}{4}\right)\right]\Big\}\,.
\label{51}
\end{split}
\end{equation}
The reason for suggesting this alternative formula is, as we will see later,
that it is more suitable for the NLIE approach. On the other hand, there is a
problem with (\ref{51}) since as can be seen from (\ref{Y10poles}) and
(\ref{Y20poles}) there are zeroes/poles dangerously close to the integration
contours. These singularities tend to the integration contours in the
$L\to\infty$ limit and asymptotically coincide. For this reason the energy
formula (\ref{51}) should be applied with care. First of all the equivalence
of it with the established formula (\ref{ener}) should be proved. In
Ref. \cite{Kazakov} the starting point for the proof of this equivalence is the
energy formula
\begin{equation}
\begin{split}
E^\prime_{\rm KL}=\frac{M}{2\pi}\int_{-\infty}^\infty{\rm d}\theta\Big\{
&\sinh v\left(\theta-\frac{i}{4}\right)\frac{{\rm d}}{{\rm d}\theta}
\ln\left[1+Y_{1,0}\left(\theta-\frac{i}{4}\right)\right]\\  
+&\sinh v\left(\theta+\frac{i}{4}\right)\frac{{\rm d}}{{\rm d}\theta}
\ln\left[1+Y_{2,0}\left(\theta+\frac{i}{4}\right)\right]\Big\}\,,
\label{51p}
\end{split}
\end{equation}
which is what one obtains from (\ref{51}) by formal partial integration. The
strategy of the proof is to shift the integration contour to the real line in
order to match the integral with the integral part of (\ref{ener}). During
this deformation of the contours we encounter singularities at the poles/zeroes
given by (\ref{Y10poles}) and (\ref{Y20poles}), provided they lie inside the
strip bordered by the contours. There are two cases. If
\begin{equation}
{\rm Im}\,\theta_{1,j}>0\qquad{\rm (case\ I)}\,,
\end{equation}
then the corresponding zeroes are inside the strip (poles are outside). If
\begin{equation}
{\rm Im}\,\theta_{1,j}<0\qquad{\rm (case\ II)}\,,
\end{equation}
then the corresponding zeroes are outside the strip (poles are inside). One can
show that during the contour deformation using Cauchy's theorem we pick up
residue contributions that make the formulas (\ref{51p}) and (\ref{ener})
exactly coincide in case I only. In case II (\ref{51p}) and (\ref{ener})
are definitely different. Using the asymptotic result (\ref{N1y1}) 
we can see that the 1-particle states belong to case I only
for even quantum numbers $n_1$. (\ref{N2y12}) shows that the lowest energy
${\cal N}=2$ symmetric state also belongs to case I. Thus luckily the states we
are interested in (lowest energy 1 and 2-particle states) are all case I states.

Another problem is that (except for the ground state) (\ref{51}) is not equal
to its formally partially integrated version (\ref{51p}). During partial
integration the boundary terms at infinity of course vanish since
$Y_{a,0}(\theta)$ are exponentially small there. However, there are boundary
terms coming from certain points on the contour. As we have seen there
are (at least for large $L$) poles and zeroes very close to each other and to
the integration contour. This implies that there must be some real 
$\hat\theta$ such that $1+Y_{1,0}(\hat\theta-i/4)$ is real and negative. Using
the standard definition of the log function, there are extra boundary
contributions coming from the fact that $\ln[1+Y_{1,0}(\theta-i/4)]$ jumps at
$\theta=\hat\theta$ by $\pm2i\pi$. Simultaneously $\ln[1+Y_{2,0}(\theta+i/4)]$
jumps at the same point by $\mp2i\pi$ since it is the complex conjugate.
For example for the standing 1-particle state we find that as $\theta$ goes
along the integration contour from $-\infty$ to $\infty$ the value of
$1+Y_{1,0}(\theta-i/4)$, starting from $+1$, crosses the negative real axis from
above at $\theta=\hat\theta$ and than goes back to $+1$ in the lower half plane.
The corresponding extra term is
\begin{equation}
-iM\sinh\left(v\left[\hat\theta-\frac{i}{4}\right]\right)\,.
\end{equation}
$\hat\theta$ is defined by
\begin{equation}
{\rm Im}\,Y_{1,0}\left(\hat\theta-\frac{i}{4}\right)=0,\qquad\quad 
{\rm Re}\,Y_{1,0}\left(\hat\theta-\frac{i}{4}\right)<-1\,.  
\label{hat1}
\end{equation}
Then also
\begin{equation}
{\rm Im}\,Y_{2,0}\left(\hat\theta+\frac{i}{4}\right)=0,\qquad\quad 
{\rm Re}\,Y_{2,0}\left(\hat\theta+\frac{i}{4}\right)<-1  
\label{hat2}
\end{equation}
giving the contribution
\begin{equation}
iM\sinh\left(v\left[\hat\theta+\frac{i}{4}\right]\right)\,.
\end{equation}
We find that by parity symmetry $\hat\theta=0$ and so
\begin{equation}
E_{{\rm KL}}=E^\prime_{\rm KL}-2M\sin\frac{\pi}{6}=E^\prime_{\rm KL}-M=E-M\,.  
\end{equation}

Similarly for our ${\cal N}=2$ state because of parity symmetry (\ref{hat1})
and (\ref{hat2}) are satisfied at
\begin{equation}
\hat\theta=\pm B_2
\end{equation}
and we have
\begin{equation}
E=E^\prime_{\rm KL}=E_{\rm KL}+2M\cosh(vB_2)\,.
\end{equation}

To summarize, we can use the energy formula (\ref{51}) for the zero momentum
0,1, and 2-particle states with the standard definition of the log function
but the correct energy is given by
\begin{equation}
E=E_{\rm KL}+{\cal N}M\cosh(vB_{\cal N})
\end{equation}
with $B_1=0$ and $B_2$ determined from the requirements
\begin{equation}
{\rm Im}\,Y_{1,0}\left(B_2-\frac{i}{4}\right)=0,\qquad\quad 
{\rm Re}\,Y_{1,0}\left(B_2-\frac{i}{4}\right)<-1\,.  
\label{B2}
\end{equation}
We emphasize that for states belonging to case II (for example moving 1-particle
states with odd momentum quantum numbers) further modifications are necessary.

\subsection{NLIE integral equations}

The unknowns to be determined are two imaginary functions along the real axis, $f_2(\eta)$ and
$f_3(\eta)$, and $\cal N$ real\footnote{The reality conditions on $f_{2,3}$ and $\beta_\alpha$
are sufficient to ensure the conjugacy properties (\ref{conj12}) of the Y-functions built out of them.}
parameters $\beta_\alpha$, $\alpha=1,\dots,{\cal N}$.

First we build the 4 Q-functions defined by
\begin{equation}
\begin{split}
q_2(\theta)&=\theta+F_2(\theta),\qquad {\rm Im}\,\theta<0,\\    
\bar q_2(\theta)&=\theta+\bar F_2(\theta),\qquad {\rm Im}\,\theta>0,\\    
q_3(\theta)&=P(\theta)+F_3(\theta),\qquad {\rm Im}\,\theta<0,\\    
\bar q_3(\theta)&=P(\theta)+\bar F_3(\theta),\qquad {\rm Im}\,\theta>0\,,    
\end{split}
\end{equation}
where
\begin{equation}
F_j(\theta)=\frac{1}{2\pi i}\int_{-\infty}^\infty\frac{f_j(\eta)}{\theta-\eta}
{\rm d}\eta,\qquad{\rm Im}\,\theta<0\,,
\end{equation}

\begin{equation}
\bar F_j(\theta)=\frac{1}{2\pi i}\int_{-\infty}^\infty\frac{f_j(\eta)}{\theta-\eta}
{\rm d}\eta,\qquad{\rm Im}\,\theta>0\,,
\end{equation}
and $P(\theta)$ is a polynomial of degree ${\cal N}+2$
\begin{equation}
P(\theta)=\sum_{j=0}^{{\cal N}+2}p_j\theta^j\,,
\end{equation}
satisfying
\begin{equation}
2P(\theta)-P(\theta-i)-P(\theta+i)=\delta_{{\cal N},0}+
(1-\delta_{{\cal N},0})\prod_{\alpha=1}^{\cal N}(\theta-\beta_\alpha)\,.
\end{equation}
We further restrict $P(\theta)$ by requiring $p_1=p_0=0$.
The absence of linear and constant terms is a kind of gauge choice.

In particular, for ${\cal N}=0$:
\begin{equation}
p_2=\frac{1}{2}.
\end{equation}
For ${\cal N}=1$:
\begin{equation}
p_3=\frac{1}{6},\qquad p_2=-\frac{1}{2}\beta_1\,.
\end{equation}
For ${\cal N}=2$:
\begin{equation}
p_4=\frac{1}{12},\qquad p_3=-\frac{1}{6}(\beta_1+\beta_2),\qquad
p_2=\frac{1}{12}+\frac{1}{2}\beta_1\beta_2\,.
\end{equation}

The functions $q_j$, $\bar q_j$ are only needed in their respective domains of definition,
except for real $\theta$ where we can use the ${\cal P}+i\pi\delta$ prescription.
In particular for $\theta$ real
\begin{equation} 
\frac{1}{2}[q_2(\theta)+\bar q_2(\theta)]=\theta+\frac{1}{2i}H(f_2)(\theta)\,,
\end{equation}
\begin{equation} 
\frac{1}{2}[q_3(\theta)+\bar q_3(\theta)]=P(\theta)
+\frac{1}{2i}H(f_3)(\theta)\,,
\end{equation}
\begin{equation}
q_j(\theta)-\bar q_j(\theta)=f_j(\theta)\,,
\end{equation}
where $H(f)$ denotes the Hilbert transform of $f$,
\begin{equation}
H(f)(x)=\frac{1}{\pi}{\cal P} \int_{-\infty}^\infty\frac{f(y)}{x-y} {\rm d}y\,.
\end{equation}

\subsubsection{The integral equations}

We introduce the notation
\begin{equation}
f^{[\pm k]}(\theta)=f\left(\theta\pm\frac{i}{2}k\right),\qquad\quad
f^\pm=f^{[\pm1]}\,.  
\end{equation}
The two NLIE equations are given by Eq.~(82) of Ref.~\cite{Kazakov} with $Z=1$. They can be written in the
form
\begin{equation}
\begin{split}  
Af_2+Bf_3&=D_1,\\
-\bar Af_2-\bar Bf_3&=D_2,
\end{split}
\end{equation}
where
\begin{equation}
\begin{split}  
A&=-iq_3^{[-2]}+\frac{i}{2}(q_3+\bar q_3),\\
B&=\phantom{-}iq_2^{[-2]}-\frac{i}{2}(q_2+\bar q_2),\\
\bar A&=\phantom{-}i\bar q_3^{[2]}-\frac{i}{2}(q_3+\bar q_3),\\
\bar B&=-i\bar q_2^{[2]}+\frac{i}{2}(q_2+\bar q_2)\,.
\end{split}
\end{equation}
The above formulas are obtained from Eqs.~ (75), (49) of Ref.~\cite{Kazakov}.
$D_1$ and $D_2$ are shorthand for the complicated expressions of the right hand side 
of Eq.~(82) of Ref.~\cite{Kazakov} with $Z=1$ and they will be given explicitly below.
We can express $f_2$ and $f_3$ from the NLIE equations. Defining
\begin{equation}
D_3= A\bar B-\bar AB
\end{equation}
we have 
\begin{equation}
\begin{split}  
f_2&=\frac{BD_2+\bar BD_1}{D_3},\\
f_3&=-\frac{AD_2+\bar AD_1}{D_3}\,.
\end{split}
\label{NLIE}
\end{equation}

\subsubsection{Exact Bethe equations}

$f_2$ and $f_3$ must be smooth functions. For ${\cal N}>0$ define the exact Bethe roots 
$\hat\theta_\alpha$ by
\begin{equation}
D_3(\hat\theta_\alpha)=0\,.
\end{equation}

Smoothness requires
\begin{equation}
\left(BD_2+\bar BD_1\right)\Big\vert_{\theta=\hat\theta_\alpha}=0\,.
\label{EBE}
\end{equation}
(Then the other numerator also vanishes.)

\subsubsection{Explicit expressions for $D_1$ and $D_2$}

\begin{equation}
D_1=\exp\left[-z\cosh\left\{v\left(\theta-\frac{i}{4}\right)\right\}\right]
T_{1,1}^{[-1/2]}{\cal T}D_{1a}D_{1b}\,,  
\end{equation}
where
\begin{equation}
{\cal T}=\frac{T_{0,0}^{[-1/2]}\,T_{3,0}^{[1/2]}}{T_{0,0}^{[-5/2]}\,
T_{3,0}^{[5/2]}}\,,
\end{equation}
and
\begin{equation}
D_{1a}=\left(\frac{T_{0,0}^{[-9/2]}}{T_{0,0}^{[-5/2]}}\right)^{*K_3},
\qquad\quad D_{1b}=\left(\frac{T_{3,0}^{[11/2]}}{T_{3,0}^{[3/2]}}\right)^{*K^{[-1]}_3}.
\end{equation}
Here the notation is
\begin{equation}
f^{*K}=\exp[\ln(f)*K],  
\end{equation}
where $*$ denotes convolution and the kernel $K_3$ is given by
\begin{equation}
K_3(\theta)=\frac{1}{\sqrt{3}[2\cosh(v\theta)+1]}\,.  
\end{equation}

\begin{equation}
D_2=\exp\left[-z\cosh\left\{v\left(\theta+\frac{i}{4}\right)\right\}\right]
T_{2,1}^{[1/2]}{\cal T}D_{2a}D_{2b}\,,  
\end{equation}
where
\begin{equation}
D_{2a}=\left(\frac{T_{3,0}^{[9/2]}}{T_{3,0}^{[5/2]}}\right)^{*K_3},
\qquad\quad D_{2b}=\left(\frac{T_{0,0}^{[-11/2]}}{T_{0,0}^{[-3/2]}}\right)^{*K^{[1]}_3}.
\end{equation}

\subsubsection{T-functions}

The building blocks for $D_1$, $D_2$ are the T-functions given by Eq.~(49) of Ref.~\cite{Kazakov}:
\begin{equation}
T_{a,s}=-i{\rm Det}{\cal M}_{a,s}\,. 
\end{equation}
(The $T_{a,s}$ expressions below are the $T^{(R)}_{a,s}$ of the R-gauge \cite{Kazakov}.)
We only need a few special cases. 
\begin{equation}
{\cal M}_{1,1}=\begin{bmatrix}
1&1&1\\
\bar q_2^{[5/2]}& q_2^{[-3/2]}&q_2^{[-7/2]}\\
\bar q_3^{[5/2]}& q_3^{[-3/2]}&q_3^{[-7/2]}
\end{bmatrix}.
\end{equation}
From this we can see that $T_{1,1}^{[-1/2]}$ is built from $\bar q_j^{[2]}$, $q_j^{[-2]}$
and $q_j^{[-4]}$, all of them in their respective domains of definition.

Similarly
\begin{equation}
{\cal M}_{2,1}=\begin{bmatrix}
1&1&1\\
\bar q_2^{[7/2]}& \bar q_2^{[3/2]}&q_2^{[-5/2]}\\
\bar q_3^{[7/2]}& \bar q_3^{[3/2]}&q_3^{[-5/2]}
\end{bmatrix}.
\end{equation}
From this we see that $T_{2,1}^{[1/2]}$ is built from $\bar q_j^{[4]}$, $\bar q_j^{[2]}$
and $q_j^{[-2]}$, again all of them in their respective domains of definition.

The other two matrices we need to construct $D_1$, $D_2$ are
\begin{equation}
{\cal M}_{0,0}=\begin{bmatrix}
1&1&1\\
q_2^{[1/2]}& q_2^{[-3/2]}&q_2^{[-7/2]}\\
q_3^{[1/2]}& q_3^{[-3/2]}&q_3^{[-7/2]}
\end{bmatrix},
\end{equation}
\begin{equation}
{\cal M}_{3,0}=\begin{bmatrix}
1&1&1\\
\bar q_2^{[7/2]}& \bar q_2^{[3/2]}&\bar q_2^{[-1/2]}\\
\bar q_3^{[7/2]}& \bar q_3^{[3/2]}&\bar q_3^{[-1/2]}
\end{bmatrix}.
\end{equation}
$T_{0,0}^{[\sigma]}$ is needed for $\sigma=-1/2,-3/2,-5/2,-9/2,-11/2$. We see all arguments
of $q_j(\theta)$ are in the lower half-plane or on the real axis. 
Similarly we only need $T_{3,0}^{[\sigma]}$ for $\sigma=1/2,3/2,5/2,9/2,11/2$.
All arguments of $\bar q_j(\theta)$ are in the upper half-plane or on the real axis.

To construct $1+Y_{1,0}^{[-1/2]}$ and $1+Y_{2,0}^{[1/2]}$ used in the energy formula (\ref{51})
we also need $T_{1,0}$ and $T_{2,0}$, i.e.
\begin{equation}
{\cal M}_{1,0}=\begin{bmatrix}
1&1&1\\
\bar q_2^{[3/2]}& q_2^{[-1/2]}&q_2^{[-5/2]}\\
\bar q_3^{[3/2]}& q_3^{[-1/2]}&q_3^{[-5/2]}
\end{bmatrix},
\end{equation}
\begin{equation}
{\cal M}_{2,0}=\begin{bmatrix}
1&1&1\\
\bar q_2^{[5/2]}& \bar q_2^{[1/2]}&q_2^{[-3/2]}\\
\bar q_3^{[5/2]}& \bar q_3^{[1/2]}&q_3^{[-3/2]}
\end{bmatrix}.
\end{equation}
Again, we only need $q_j^{[\sigma]}$ for $\sigma=0,-2,-4$ and  $\bar q_j^{[\sigma]}$ for
$\sigma=0,2,4$, all functions in their respective domains of definition or on the real axis.
This is why we insist on using the energy formula (\ref{51}). The energy formula (\ref{ener}) is
numerically more stable, but requires analytical continuation of the Q-functions.

\subsubsection{Iteration}

The NLIE equations are usually solved iteratively. Start from some approximation
\begin{equation}
\{f^{(\nu)}_j(\eta)\},\ (j=2,3);\qquad  \{\beta^{(\nu-1)}_\alpha\},\ \alpha=1,\dots,{\cal N}.
\end{equation}

To prepare the iteration we compute the approximations
\begin{equation}
D^{(\nu)}_j, \ (j=1,2,3), \qquad A^{(\nu)}, B^{(\nu)}, \bar A^{(\nu)}, \bar B^{(\nu)}.
\label{inter}
\end{equation}
In these computations we use 
\begin{equation}
\{f^{(\nu)}_j(\eta)\},\ (j=2,3);
\qquad  \{\beta^{(\nu)}_\alpha\},\ \alpha=1,\dots,{\cal N}.
\end{equation}
Note we do not yet know the $\beta^{(\nu)}_\alpha$ so these are just
intermediate parameters in the computation of (\ref{inter}).

Also compute the approximate Bethe roots $\hat\theta^{(\nu)}_\alpha$ by solving
\begin{equation}
D_3^{(\nu)}(\hat\theta_\alpha^{(\nu)})=0.
\end{equation}
They depend on the yet unknown parameters $\beta^{(\nu)}_\alpha$.

These parameters are now determined by the approximation to the exact Bethe equations
\begin{equation}
\left(B^{(\nu)}D^{(\nu)}_2+\bar B^{(\nu)}D^{(\nu)}_1\right)\Big\vert_{\theta=\hat\theta^{(\nu)}_\alpha}=0.
\end{equation}
Now we know the $\beta^{(\nu)}_\alpha$, we can compute the next approximation to the spectral densities
\begin{equation}
\begin{split}  
f^{(\nu+1)}_2&=\frac{B^{(\nu)}D^{(\nu)}_2+\bar B^{(\nu)}D^{(\nu)}_1}{D_3^{(\nu)}},\\
f^{(\nu+1)}_3&=-\frac{A^{(\nu)}D^{(\nu)}_2+\bar A^{(\nu)}D^{(\nu)}_1}{D_3^{(\nu)}}.
\end{split}
\end{equation}

The zeroth approximation is
\begin{equation}
f_2^{(0)}=f_3^{(0)}=0
\end{equation}
and in this case
\begin{equation}
\hat\theta^{(0)}_\alpha=\beta^{(0)}_\alpha,\quad \alpha=1,\dots,{\cal N}.
\end{equation}
As shown in subsection 7.4 of Ref.~\cite{Kazakov} the zeroth approximation to the exact Bethe equations
reduces to the asymptotic Bethe equations
\begin{equation}
{\rm e}^{iz\sinh\left(v\hat\theta^{(0)}_\alpha\right)}\,
\prod_{\beta=1}^{\cal N} S\left(v(\hat\theta^{(0)}_\alpha-\hat\theta_\beta^{(0)})\right)=-1.
\end{equation}

\subsection{Numerical results}

We have calculated the energies for the SU$(3)$ states $r_p=(p,0,0)$ for $p=0,1,2$
(vacuum, standing 1-particle and parity symmetric 2-particle states) to 12 digits numerically for small volumes.
These energies are denoted by ${\cal E}_p$ in Table \ref{tab:table2}.
\begin{table}[ht]
  \centering
  \caption{SU(3) energies for $p=0,1,2$.
    The last column presents values of a 2-particle NLIE parameter
    $\beta_1$}
  \label{tab:table2} 
  \vspace{0.5cm}
  \begin{tabular}{|l|l|l|l|l|}
    \hline
    \quad$z$ & $\quad L\mathcal{E}_0$ & $\quad L\mathcal{E}_1$
    & $\quad L\mathcal{E}_2$ & $\quad \beta_1$ \\
    \hline
    0.01& $-3.420577098958$&$-2.76292779616$&$-1.777099(3) $&$1.4048256851$ \\
    0.02& $-3.342476728778$&$-2.62162678936$&$-1.5412988(6)$&$1.2641055097$ \\
    0.03& $-3.288368668518$&$-2.52428616251$&$-1.379383(3) $&$1.1817120504$ \\
    0.04& $-3.245217335083$&$-2.44700634691$&$-1.251156(4) $&$1.12322793788$\\
    0.05& $-3.208524192169$&$-2.38154930287$&$-1.14279(1)  $&$1.07785551177$\\
    0.06& $-3.176149111708$&$-2.32399857099$&$-1.0477086(6)$&$1.04078170872$\\
    0.07& $-3.146890539710$&$-2.27215624177$&$-0.9622193(7)$&$1.00943765385$\\
    0.08& $-3.120000348735$&$-2.22465417518$&$-0.8840251(2)$&$0.98228925202$\\
    0.09& $-3.094978291235$&$-2.18057765731$&$-0.811592(3) $&$0.95834659282$\\
    0.1&  $-3.071471895079$&$-2.13928219543$&$-0.743839(1) $&$0.93693354156$\\
    0.2&  $-2.883458484829$&$-1.81286374544$&$-0.2121368(2)$&$0.79626871125$\\
%    0.3&  $-2.736375113643$&$$ & $$ &$$\\
%    0.4&  $-2.607723094037$&$$ & $$ &$$\\
    0.5&  $-2.489784492441$&$-1.14837251407$&$\phantom{-}0.849784(1)$&$0.611803404$\\
%    0.6&  $-2.378973211746$&$$ & $$ &$$\\
%    0.7&  $-2.273355006231$&$$ & $$ &$$\\
%    0.8&  $-2.171794376719$&$$ & $$ &$$\\
%    0.9&  $-2.073591799276$&$$ & $$ &$$\\
    1.0&  $-1.97830660727$ &$-0.29936999966$&$\phantom{-}2.180019(2)$&$0.4754496334$\\
    \hline
  \end{tabular}
\end{table}
The results agree with those of Ref. \cite{Kazakov} up to the numerical precision given there.

Studying the volume dependence of the 2-particle $\beta_1$ parameter for small volumes
we conjecture that it is given by a perturbative expansion in the running coupling $\alpha_{\rm J}$
(see subsection 5.1) with coefficients $a_1,a_2,a_3,\dots$
\begin{equation}
\beta_1=\frac{a_1}{\alpha_{\rm J}}+a_2+a_3\alpha_{\rm J}
+{\rm O}(\alpha_{\rm J}^2)\,,  
\end{equation}
where
\begin{equation}
a_1\simeq\frac{\pi}{48}\,.
\end{equation}

% K&L
%    $10^{-2}$& $-3.4206$  &$-2.7629$ & $-1.777$ \\
%    $10^{-1}$& $-3.0715$  &$-2.1393$ & $-0.7439$ \\
%    $10^{0}$ & $-1.9783$  &$-0.2993$ & $\phantom{-}2.180$ \\
%    $10^{1}$ & $-0.0010683$  &$9.995$ & $\phantom{-}20.66$ \\

\end{appendix}

\end{document}